\documentclass[10pt,aps,prb,twocolumn,preprintnumbers,amsmath,amssymb,floatfix,showpacs,citeautoscript]{revtex4-2}
\usepackage[latin1]{inputenc}
\usepackage[T1]{fontenc}
\usepackage[english]{babel}
\usepackage[pdftex]{graphicx}
\usepackage{amsmath}
\usepackage{times}
\usepackage[usenames,dvipsnames,svgnames,table]{xcolor}
\usepackage[colorlinks,plainpages=false,linkcolor=blue,urlcolor=blue,citecolor=blue,pdfpagemode=UseNone]{hyperref}
\usepackage{hyperref}

\renewcommand{\AA}{\text{\r{A}}}

\newcommand{\alphaR}{\alpha_\text{R}}
\newcommand{\alphaRRR}{\tilde{\alpha}_\text{R}}
\newcommand{\dzb}{\Delta z_B}

\begin{document}

\title
{
\boldmath
Rashba spin-orbit coupling in infinite-layer nickelate films on SrTiO$_3$(001) and KTaO$_3$(001) %
}

\author{Benjamin Geisler}
\email{benjamin.geisler@ufl.edu}
\affiliation{Department of Physics, University of Florida, Gainesville, Florida 32611, USA}
\affiliation{Department of Materials Science and Engineering, University of Florida, Gainesville, Florida 32611, USA}

\date{\today}

\begin{abstract}
The impact of spin-orbit interactions in NdNiO$_2$/SrTiO$_3$(001) and NdNiO$_2$/KTaO$_3$(001) %
is explored by performing density functional theory simulations including a Coulomb repulsion term.
Polarity mismatch drives the emergence of an interfacial two-dimensional electron gas %
in NdNiO$_2$/KTaO$_3$(001) involving the occupation of Ta $5d$ conduction-band states, %
which is twice as pronounced as in NdNiO$_2$/SrTiO$_3$(001). %
We identify a significant anisotropic $k^3$ Rashba spin splitting of the respective $d_{xy}$ states in both systems
that results from the broken inversion symmetry at the nickelate-substrate interface %
and exceeds the width of the superconducting gap.
In NdNiO$_2$/KTaO$_3$(001), the splitting reaches $210$~meV, which is comparable to Bi(111) surface states.
At the surface, the Ni $3d_{x^2-y^2}$-derived states exhibit a linear Rashba effect with %
$\alphaR \sim 125$~meV~$\AA$, exemplifying its orbital selectivity.
The corresponding Fermi sheets present a reconstructed circular shape due to the electrostatic doping,
but undergo a Lifshitz transition towards a cuprate-like topology deeper in the film %
that coincides with a realignment of their spin texture.
These results promote surface and interface polarity as interesting design parameters to control spin-orbit physics in infinite-layer nickelate heterostructures.
\end{abstract}

\maketitle

Spin-orbit coupling (SOC) correlates the electron spin and the crystal lattice and %
gives rise to a multitude of intriguing quantum phenomena. %
A prominent example is the Rashba effect~\cite{BychkovRashba:84},
which originates from the broken inversion symmetry at interfaces and surfaces
and is of fundamental relevance in spintronics,
as it permits the electric control of the spin precession~\cite{SpintronicsFabian:04}. %
Moreover, it plays a key role in the quest to realize Majorana zero modes
by exploiting the proximity effect between a material with strong SOC and a superconductor~\cite{Lutchyn:10, Oreg:10, Lutchyn:18}.
The pronounced coupling of lattice, orbital, spin, and charge degrees of freedom
in transition metal oxides~\cite{OxideRoadmap:16}
renders them a particularly interesting platform %
to explore superconductivity in conjunction with a two-dimensional Rashba system.
Paradigmatic is the LaAlO$_3$/SrTiO$_3$(001) interface~\cite{Ohtomo:2004, Nakagawa:06, Thiel:06, PentchevaPickett:09, Bell:09},
where the Rashba effect is considered to stabilize the emergent superconducting phase~\cite{Reyren:07, LAOSTO-Rashba-CavigliaTriscone:10, BenShalom:10}
and may reconcile superconductivity and magnetism~\cite{Dikin:11, Michaeli:12}.

The very recent observation of superconductivity in Sr-doped NdNiO$_2$, PrNiO$_2$, and LaNiO$_2$ films grown on 
SrTiO$_3$(001) (STO)~\cite{Li-Supercond-Inf-NNO-STO:19, Li-Supercond-Dome-Inf-NNO-STO:20, Osada-PrNiO2-SC:20, Zeng-Inf-NNO:20}
sparked considerable interest in $3d^9$ infinite-layer ($AB$O$_2$) nickelates,
which are formally isoelectronic to the cuprates~\cite{Nomura-Inf-NNO:19, JiangZhong-InfNickelates:19, Sakakibara:20, JiangBerciuSawatzky:19, Botana-Inf-Nickelates:19, Lechermann-Inf:20, Si-Zhonh-Held:InfNNO-Hydrogen:20, NNO-SC-Thomale:20, Gu-NNO2:20, Lu-MagExNdNiO2:21, Ortiz-NNO:21, Lechermann-Inf:21, SahinovicGeisler:21, Wang-IL-Pauli:21, Optical-IL-Kotliar:21, Zeng-Inf-NNO:22, KreiselLechermann-IL:22, Rossi-IL-CO:22, Fowlie-IL-IntrinsicMag:22, SahinovicGeisler:22, Wang-Pressure-PNO:22, Pan-ILSC:22, GoodgeGeisler-NNO-IF:22}.
However, superconductivity remained elusive in the respective bulk compounds so far~\cite{Li-NoSCinBulkDopedNNO:19, Wang-NoSCinBulkDopedNNO:20}.
Early work unraveled the key role of the polar interface~\cite{GeislerPentcheva-InfNNO:20, BernardiniCano:20, He-IL:20, Zhang-IL:20, GeislerPentcheva-NNOCCOSTO:21}:
The polarity mismatch drives the formation of a two-dimensional electron gas (2DEG) in the substrate
by occupation of the interfacial Ti~$3d$ states~\cite{GeislerPentcheva-InfNNO:20, GeislerPentcheva-NNOCCOSTO:21} %
that is considerably more pronounced than in LaAlO$_3$/STO(001)~\cite{PentchevaPickett:09}.
Therefore, these artificial oxides naturally host a 2DEG proximate to a novel class of superconductor.

An intriguing alternative to STO is KTaO$_3$ (KTO),
which exhibits stronger spin-orbit interactions~\cite{KTaO3-SOC-Hwang:12}
and attracts increasing attention
due to the emergence of superconductivity at different (111) and (110) heterointerfaces~\cite{Liu-SC-KTO-111-IFs:21, Chen-SC-LAOKTO110:21, Gupta-KTO-Review:22}.
Moreover, 2DEGs have been reported in a variety of (001)-oriented perovskite systems~\cite{Zou-LTOKTO-001:15, Zhang-amorphLAOKTO-001:19, Wadehra-LVOKTO001:20}.

Here we explore the impact of spin-orbit interactions in NdNiO$_2$/STO(001) and NdNiO$_2$/KTO(001)
by performing density functional theory simulations including a Coulomb repulsion term in explicit film geometry.
The distinct polar discontinuity at the Nd$^{3+}$/(TiO$_2$)$^{0}$ versus Nd$^{3+}$/(TaO$_2$)$^{1+}$ interface
results in the formation of a twice as pronounced 2DEG in the KTO system %
by occupation of Ta $5d$ conduction-band states, predominantly $d_{xy}$,
which is accompanied by ferroelectric-like displacements of the Ta ions in the substrate. %
We find that the broken inversion symmetry at the surface and the interface gives rise to an anisotropic Rashba spin splitting %
that exceeds the size of the superconducting gap.
Surprisingly, it reaches values at the KTO interface that are comparable to Bi(111) surface states. %
The Ti and Ta $d_{xy}$ orbitals in the substrate exhibit a fundamentally different response to SOC than the Ni $3d_{x^2-y^2}$ orbitals in the film, %
i.e., a predominantly cubic versus linear momentum dependence of the spin splitting.
We observe a substantial reconstruction of the Ni $3d_{x^2-y^2}$-derived Fermi sheets at the surface due to the electrostatic doping,
suggesting local modifications of the pairing mechanism.
Layer-wise tracking of these states throughout the nickelate film unveils a Lifshitz transition towards a cuprate-like topology
that coincides with a reorientation of their spin texture.
These insights exemplify the impact of surface and interface polarity on spin-orbit-driven phenomena in infinite-layer nickelate heterostructures. %

\begin{figure*}
\begin{center}
\includegraphics[width=\textwidth]{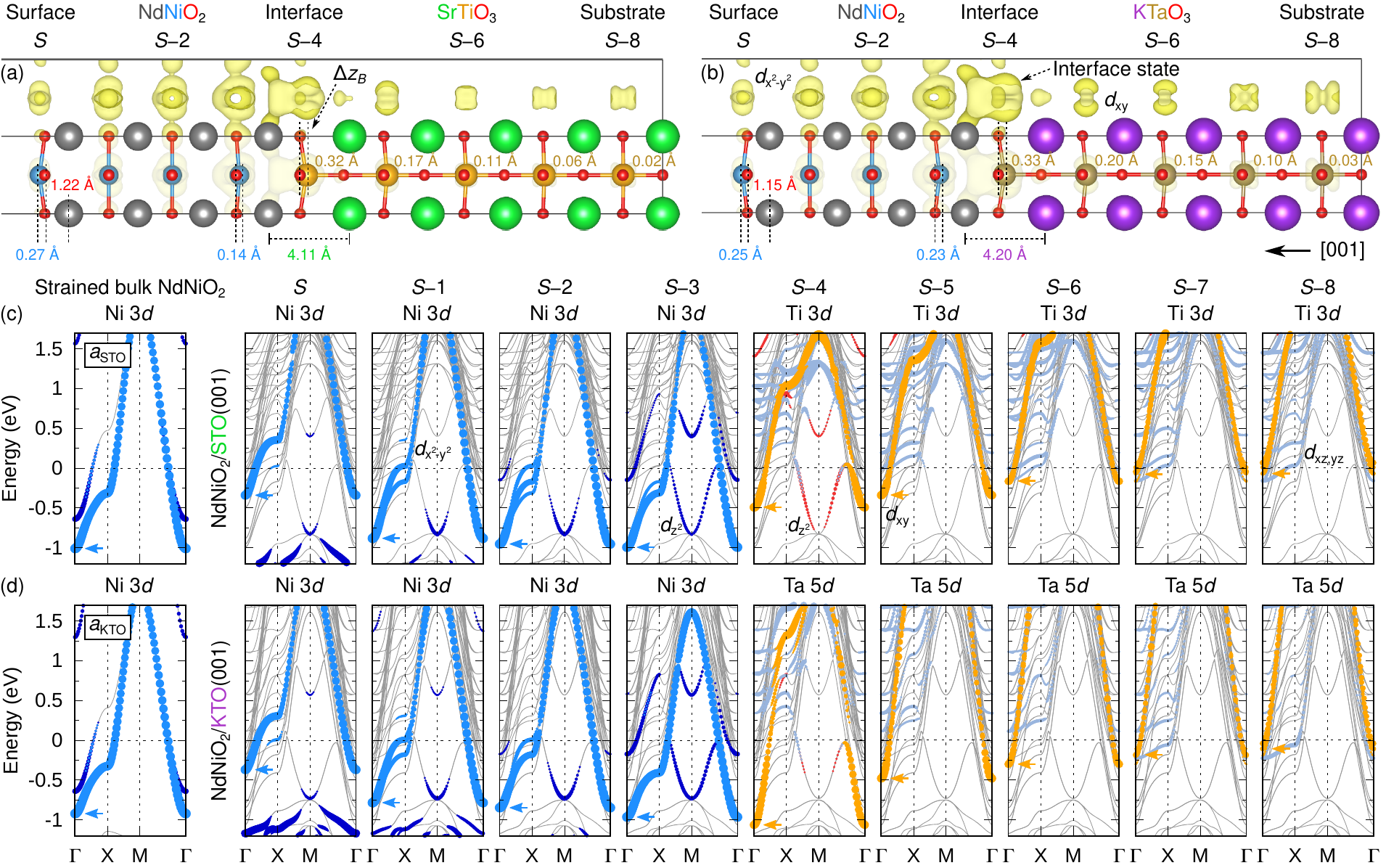}
\caption{\label{fig:LayerBands} Optimized geometry of (a) (NdNiO$_2$)$_4$/STO(001)~\cite{GeislerPentcheva-InfNNO:20} and (b) (NdNiO$_2$)$_4$/KTO(001), together with the electron density integrated between $-0.75$~eV and the Fermi energy.
(c,d)~The corresponding layer-resolved band structures (here from DFT$+U$ without SOC) show the polarity-driven emergence of an interfacial 2DEG in both substrates by occupation of Ti~$3d$ and Ta~$5d$ conduction-band states, respectively (orange arrows), as well as the electrostatic modulation of the Ni~$3d$ states in the polar infinite-layer nickelate film (blue arrows).
Light- and dark-blue bands depict Ni~$d_{x^2-y^2}$ and $d_{z^2}$ states; orange, gray-blue, and red bands depict Ti and Ta~$d_{xy}$, $d_{xz,yz}$, and $d_{z^2}$ states; and the thin gray lines represent the total band structure of the supercells. Results for strained bulk NdNiO$_2$ are provided as reference (left panels).
}
\end{center}
\end{figure*}

\textit{Methodology. --}
We performed first-principles simulations in the framework of density functional theory (DFT~\cite{KoSh65})
as implemented in the \textit{Vienna Ab initio Simulation Package} (VASP)~\cite{USPP-PAW:99, PAW:94}, 
employing the PBE exchange-correlation functional~\cite{PeBu96}  %
and a wave-function cutoff of $520$~eV. %
Static correlation effects were considered within the DFT$+U$ formalism~\cite{LiechtensteinAnisimov:95, Dudarev:98}
employing $U-J=3$~eV at the Ni and Ti sites %
and $U-J=1$~eV at the Ta sites,
similar to previous work~\cite{Liu-NNO:13, Botana-Inf-Nickelates:19, Guo:17, Geisler-LNOSTO:17, WrobelGeisler:18, GeislerPentcheva-LNOLAO:18, GeislerPentcheva-LNOLAO-Resonances:19, Koeksal:19, GeislerPentcheva-InfNNO:20, GeislerPentcheva-NNOCCOSTO:21, Geisler-VO-LNOLAO:22}.
Subsequent to a self-consistent calculation, %
we shift to a spinor representation, account for spin-orbit interactions, and thus obtain an updated electronic structure.

(NdNiO$_2$)$_4$/STO(001) and (NdNiO$_2$)$_4$/KTO(001) are modeled in explicit film geometry by using tetragonal supercells,
fixing the in-plane lattice parameter to
$a_\text{STO} = 3.905~\AA$ and $a_\text{KTO} = 3.988~\AA$~\cite{Gupta-KTO-Review:22},
respectively.
The supercells are symmetric, consisting in total of $9.5$ unit cells of substrate, $4+4$ unit cells of infinite-layer nickelate film, and about $20~\AA$ of vacuum region.

The Brillouin zone was sampled employing a \mbox{$16\times16\times1$} Monkhorst-Pack $\Vec{k}$-point grid~\cite{MoPa76}
in conjunction with a Gaussian smearing of $0.005$~Ry.
The ionic positions were accurately optimized, reducing ionic forces below $1$~mRy/a.u.

\textit{Ionic relaxations and electronic reconstruction. --}
Figures \ref{fig:LayerBands}(a) and~(b) show the optimized geometry %
of (NdNiO$_2$)$_4$/STO(001)~\cite{GeislerPentcheva-InfNNO:20} and (NdNiO$_2$)$_4$/KTO(001), respectively.
The apical $A$-site distances are significantly enhanced at the interface ($S-4$) to
$d_\text{Nd-Sr} = 4.11~\AA$ for STO %
and even
$d_\text{Nd-K} = 4.20~\AA$ for KTO.
This observation can be attributed to the electrostatic doping~\cite{GeislerPentcheva-InfNNO:20}
and resembles the
increased La-Sr distance at the $n$-type LaNiO$_3$/STO(001) interface ($\sim 4.06~\AA$)~\cite{Geisler-LNOSTO:17, ZhangKeimer:14, Hwang:13}.
Conversely, the distance between the surface NiO$_2$ layer ($S$) and the
subsurface Nd layer is considerably contracted to %
$d_\text{Nd-O} = 1.22~\AA$ (STO) and %
$1.15~\AA$ (KTO). %

Both systems exhibit a prominent buckling of the NiO$_2$ layers at the surface ($S$) [Figs.~\ref{fig:LayerBands}(a,b)],
the Ni ions being displaced outwards from the respective oxygen layer
by $\dzb = 0.27~\AA$ (STO) and $0.25~\AA$ (KTO).
At the interface ($S-3$), the %
Ni displacements point inwards and are more distinct,
i.e., $0.14~\AA$ (STO) and $0.23~\AA$ (KTO).
Furthermore, the Ti and Ta ions show a sizable %
inwards ferroelectric-like displacement %
of $\dzb \sim 0.3~\AA$ in layer $S-4$
that decays %
with increasing distance from the interface.
Close comparison of the two systems reveals consistently larger displacements $\dzb$ in the KTO substrate,
even though bulk KTO is cubic and paraelectric down to at least $4.2$~K~\cite{KTO-Bulk-Wemple:65, KTO-noFE:77, Gupta-KTO-Review:22}.
In summary, these results point to a substantial difference between the two materials combinations. %

The vertical ionic displacements are a sensitive response to the internal electric fields~\cite{GeislerPentcheva-InfNNO:20, GeislerPentcheva-NNOCCOSTO:21}. %
This becomes evident in Figs.~\ref{fig:LayerBands}(c) and~(d),
which compare the layer-resolved DFT$+U$ electronic structure of the two systems.
Similar to NdNiO$_2$/STO(001), %
the polar discontinuity at the NdNiO$_2$/KTO(001) interface %
drives the formation of a 2DEG due to the occupation of Ta~$5d$ conduction-band states in the substrate,
predominantly the planar $d_{xy}$ orbitals, %
which are bent down by $-1.06$~eV at the $\Gamma$ point in layer $S-4$.
This electronic reconstruction and the orbital order present at the Ti and Ta sites are clearly
visible in the electron density [Figs.~\ref{fig:LayerBands}(a,b)]. %
For comparison, the respective Ti~$3d$ states bend down by $-0.49$~eV ($-0.55$~eV in Ref.~\cite{GeislerPentcheva-InfNNO:20}).
The strongly enhanced Ta~$5d$ occupation relative to Ti~$3d$ 
can be traced back to the distinct ${3+}/{1+}$ versus ${3+}/0$ interface polarity.
The band bending decays exponentially with increasing distance from the interface,
paralleling the evolution of the ferroelectric-like displacements $\dzb$.
Hence, the potential well confining the 2DEG is considerably more asymmetric in NdNiO$_2$/KTO(001) than in NdNiO$_2$/STO(001). %
A concomitant electrostatic modulation of the Ni~$3d$ states
leads to an overall charge depletion (hole doping) in the polar films for both systems, particularly near the surface [Figs.~\ref{fig:LayerBands}(c,d)].

\begin{figure}
\begin{center}
\includegraphics[width=\linewidth]{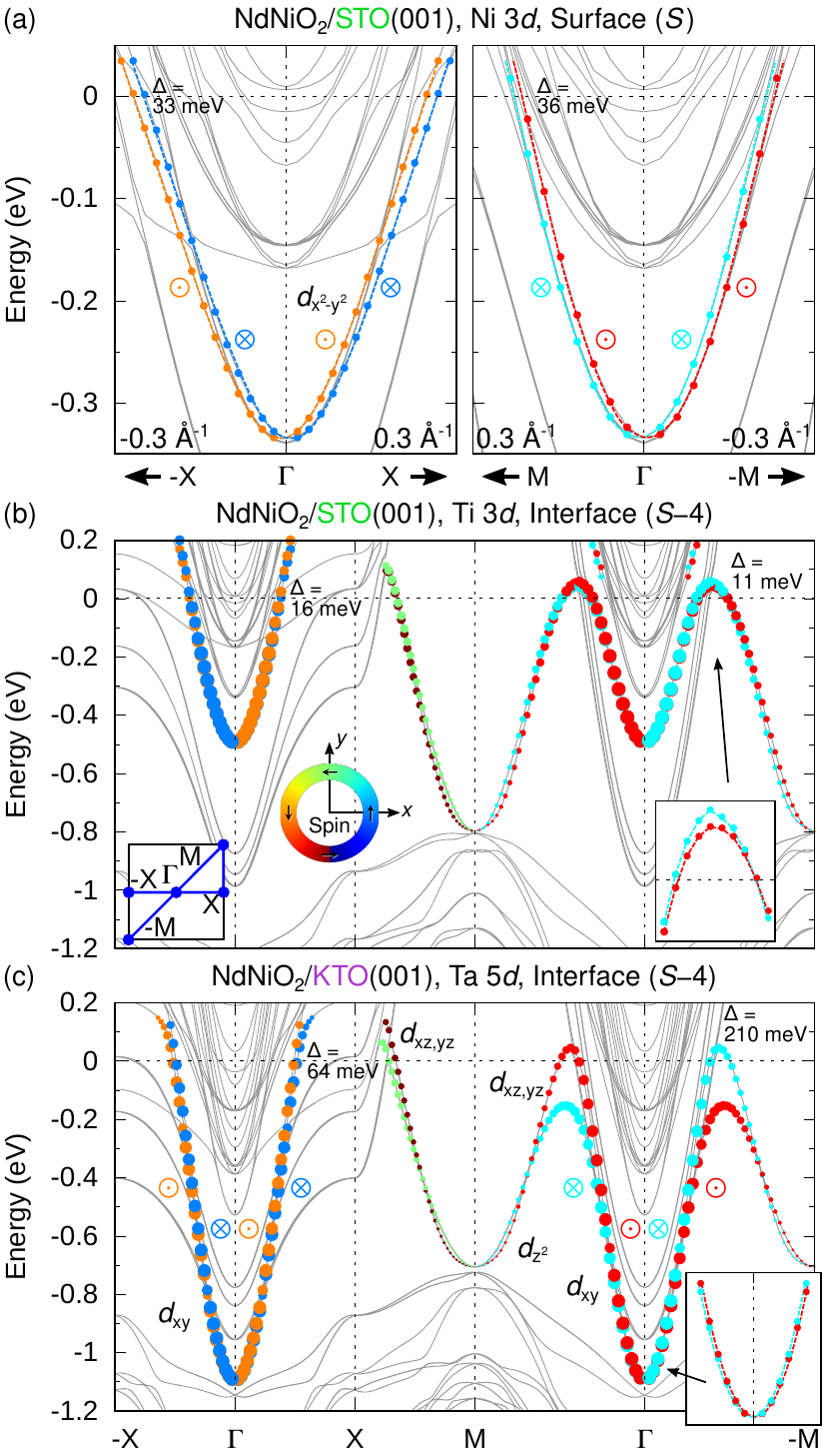}
\caption{\label{fig:BandsSOC} DFT$+U+$SOC surface and interface electronic structure in NdNiO$_2$/STO(001) and NdNiO$_2$/KTO(001). %
The colored points represent the orbital projections, %
while the color encodes the spin direction in the $xy$ plane.
The splitting of bands with opposite spin is clearly visible (e.g., blue vs.\ orange and cyan vs.\ red points). %
(a)~The Ni-$3d_{x^2-y^2}$-derived surface state ($S$), plotted exemplarily for NdNiO$_2$/STO(001) along two directions,
displays the characteristic Rashba shape around the $\Gamma$ point.
(b,c)~The interface state ($S-4$) exhibits a similar effect,
but with a massively boosted splitting in NdNiO$_2$/KTO(001). %
Intriguingly, the outer Ta~$5d$ cone %
does not cross the Fermi energy along $\Gamma$-$M$, in sharp contrast to the inner cone. %
This behavior is fundamentally distinct from NdNiO$_2$/STO(001).
\vspace{-1em}
}
\end{center}
\end{figure}

\textit{Anisotropic and orbital-selective Rashba effect. --}
We now explore the impact of SOC on the electronic structure %
at the surface of the nickelate film and in the 2DEG near the interface,
where the electronic reconstruction is most pronounced. %
The polar discontinuities result in strong electric fields
perpendicular to the basal conduction plane
that are perceived as magnetic fields by a moving electron and thus couple to its spin.
Specifically, this gives rise to the Rashba effect, which can be described by %
$H_\text{R} = \alphaR (\vec{e}_z \times \vec{k}) \cdot \vec{\sigma} ,$
where $\vec{\sigma}$ is the Pauli vector %
and $\alphaR$ is the Rashba coupling constant.
The latter is a measure of the SOC strength and depends on the local electric fields~\cite{Bihlmayer-Topo:17}.
It lifts the spin degeneracy of the bands
and manifests as a linear splitting $\Delta(k) = 2 \alphaR k$
for a free-electron 2DEG model. %
First-principles simulations provide a more comprehensive perspective %
and allow us to explore the Rashba effect layer resolved,
specifically the aspect of anisotropy, higher-order $k^3$ effects, and the distinct response of the different transition-metal $d$ orbitals. %

Figure~\ref{fig:BandsSOC}(a) shows the Ni-$3d_{x^2-y^2}$-derived surface state in layer $S$, exemplarily for NdNiO$_2$/STO(001);
we obtained similar results for NdNiO$_2$/KTO(001). %
The characteristic Rashba shape emerges around the $\Gamma$~point in the basal plane:
The quasi-parabolic band splits into two distinct cones of opposite spin
(represented here by different colors).
The spin splitting is anisotropic, reflecting the fourfold $D_{4h}$ symmetry of the structure, %
and amounts to $\Delta = 33$~meV along $\Gamma$-$X$ and $36$~meV along $\Gamma$-$M$ at the Fermi level.
This corresponds to a wave vector of $\Delta k_\parallel \sim 0.02~\AA^{-1}$,
which is surprisingly larger than $\Delta k_\parallel \sim 0.01~\AA^{-1}$ calculated for the polar KTO(001) surface~\cite{KTaO3-SOC-Hwang:12},
even though NdNiO$_2$ consists of much lighter elements and
shows negligible impact of SOC in the bulk~\cite{SahinovicGeislerPentcheva:23}.

Next, we investigate the state in the first substrate layer [$S-4$; Figs.~\ref{fig:BandsSOC}(b) and~(c)]. %
Around the $\Gamma$ point, it exhibits a quasi-parabolic dispersion, indicative of the 2DEG, and predominantly $d_{xy}$ orbital character.
In NdNiO$_2$/STO(001) [Fig.~\ref{fig:BandsSOC}(b)],
the Rashba splitting at the Fermi level amounts to $\Delta = 16$~meV along $\Gamma$-$X$ and $11$~meV along $\Gamma$-$M$, %
which is smaller than the results obtained for the surface state.
Interestingly, along $\Gamma$-$M$, the bands bend down again after initially rising to $\sim 0.05$~eV, %
concomitantly changing their predominant orbital character from Ti~$d_{xy}$ to Ti~$d_{xz,yz}$ and finally to Ni~$d_{z^2}$ at the $M$ point.
In contrast, the corresponding state in layer $S-5$ presents a consistently parabolic dispersion around the $\Gamma$ point [cf.~Figs.~\ref{fig:LayerBands}(c,d)].

For NdNiO$_2$/KTO(001) [Fig.~\ref{fig:BandsSOC}(c)], the dispersion of the interface state is qualitatively similar to NdNiO$_2$/STO(001).
However, the splitting along $\Gamma$-$X$ is considerably enhanced to $\Delta = 64$~meV,
which is four times as high as for the STO system.
This demonstrates that shifting from $Z_\text{Ti}=22$ to $Z_\text{Ta}=73$ promotes the impact of SOC substantially.
An additional contribution stems from the
stronger asymmetry of the confining potential well in NdNiO$_2$/KTO(001) owing to the distinct interface polarity [Fig.~\ref{fig:LayerBands}(d)].
Intriguingly, we observe a spin splitting of $\Delta = 210$~meV along $\Gamma$-$M$, %
which is $19$ times as high as for the STO system and
reminiscent of Bi(111) surface states ($\sim 0.2$~eV~\cite{Bi111-SOC:04}; $Z_\text{Bi}=83$). %
Even the shape of the spin-split bands %
bears striking resemblance to Bi(111).
A notable difference, however, is the suppressed contribution of the counter-clockwise Ta~$5d$ cone to the Fermi surface, %
in sharp contrast to the clockwise cone [Fig.~\ref{fig:BandsSOC}(c)].
We verified this result, which is fundamentally distinct from NdNiO$_2$/STO(001),
by performing fully self-consistent SOC calculations.
Compellingly, the pristine KTO(001) surface lacks a comparable effect %
and exhibits a much smaller spin splitting~\cite{KTaO3-SOC-Hwang:12}.

\begin{figure}
\begin{center}
\includegraphics[width=\linewidth]{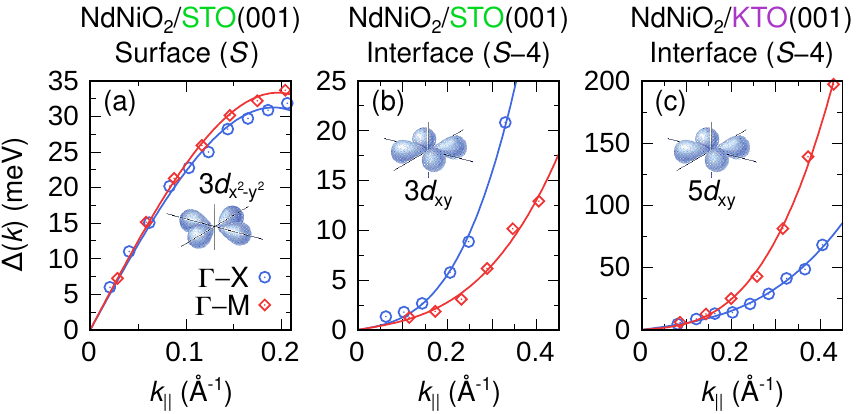}
\caption{\label{fig:Delta-k-Fit} Momentum-resolved spin splitting $\Delta(k)$ around the $\Gamma$ point
for selected layers of (a,b)~NdNiO$_2$/STO(001) and (c)~NdNiO$_2$/KTO(001) from DFT$+U+$SOC (symbols; cf.~Fig.~\ref{fig:BandsSOC}).
The lines represent fits to $\Delta(k) = 2 \alphaR k^{} + 2 \alphaRRR k^3$ (Table~\ref{tab:AlphaR}).
The qualitative differences between the three panels reflect the orbital selectivity of the anisotropic Rashba effect at the surface versus interface,
as well as the role of lighter versus heavier substrate elements.
}
\end{center}
\end{figure}

\begin{table}[b]
	\centering
	\vspace{-1.5ex}
	\caption{\label{tab:AlphaR}Rashba constants $\alphaR$ (meV~$\AA$) and $\alphaRRR$ (meV~$\AA^{3}$) obtained by fitting the DFT$+U+$SOC data for the surface and two substrate layers near the interface (cf.~Fig.~\ref{fig:Delta-k-Fit}). Additionally, the corresponding spin splitting energy $\Delta$ (meV) at the Fermi level is provided.}
	\begin{ruledtabular}
	\begin{tabular}{lccccccc}
      System & & \multicolumn{3}{c}{------ $\Gamma$-$X$ ------} & \multicolumn{3}{c}{------ $\Gamma$-$M$ ------}  \\
		 Layer & Orbital & $\Delta$ & $\alphaR$ & $\alphaRRR$ & $\Delta$ & $\alphaR$ & $\alphaRRR$  \\
		\hline
		\multicolumn{8}{l}{NdNiO$_2$/STO(001)} \\
		$S$       & Ni $3d_{x^2-y^2}$ & $33$ & $123$ & $-1116$ & $36$ & $128$ & $-1107$ \\
		$S-4$     & Ti $3d_{xy}$ & $16$ & $3.9$ & $251$ & $11$ & $3.8$ & $79$ \\
		$S-5$     & Ti $3d_{xy}$ & $8$ & $2.4$ & $278$ & $10$ & $1.5$ & $405$ \\
		\multicolumn{8}{l}{NdNiO$_2$/KTO(001)} \\
		$S$       & Ni $3d_{x^2-y^2}$ & $32$ & $119$ & $-1092$ & $35$ & $123$ & $-1052$ \\
		$S-4$     & Ta $5d_{xy}$ & $64$ & $24$ & $355$ & $210$ & $14$ & $1185$ \\
		$S-5$     & Ta $5d_{xy}$ & $15$ & $14$ & $409$ & $36$ & $11$ & $972$ \\
	\end{tabular}
	\end{ruledtabular}
\end{table}

Motivated by this observation,
we analyze the momentum-resolved spin splitting $\Delta(k)$
of the states shown in Fig.~\ref{fig:BandsSOC} along two directions
from the $\Gamma$ point to roughly the Fermi wave vector $k_\text{F}$ in Fig.~\ref{fig:Delta-k-Fit}.
The data points show a clear nonlinearity,
but can be accurately fitted to
$\Delta(k) = 2 \alphaR k^{} + 2 \alphaRRR k^3$
despite the broad $k$ range.
The three panels reveal the complexity of the Rashba effect in infinite-layer nickelate heterostructures. %
First, they underpin that $\Delta(k)$ is anisotropic and generally more pronounced along $\Gamma$-$M$. %
Surprisingly, NdNiO$_2$/STO(001) shows an inverted behavior in layer $S-4$, with a stronger splitting along $\Gamma$-$X$ [Fig.~\ref{fig:Delta-k-Fit}(b)].
Second, the $d_{x^2-y^2}$-derived states in the nickelate film exhibit a fundamentally distinct response to SOC %
than the $d_{xy}$-derived states in the substrate,
albeit both are planar.
At the surface, the linear term is very strong ($\alphaR \sim 125$~meV~$\AA$, Table~\ref{tab:AlphaR}).
Simultaneously, $\alphaRRR \sim -1.1$~eV~$\AA^{3}$ is highly negative due to a profound nonlinearity for $k>0.1~\AA^{-1}$ [Fig.~\ref{fig:Delta-k-Fit}(a)].
In sharp contrast, the substrate layers feature a dominant $k^3$ term with positive $\alphaRRR$ [Figs.~\ref{fig:Delta-k-Fit}(b,c), Table~\ref{tab:AlphaR}].
The linear term is at least one order of magnitude smaller than at the surface and decays with increasing distance to the interface.
Despite this orbital selectivity, %
$\alphaR$ is not fully quenched by symmetry, as suggested for the KTO(001) surface \cite{KTaO3-SOC-Hwang:12}.
Finally, the role of heavier versus lighter substrate elements manifests in massively enhanced Rashba constants $\alphaR$ and $\alphaRRR$ near the NdNiO$_2$/KTO(001) interface (Table~\ref{tab:AlphaR}),
in line with the boosted splittings $\Delta$ discussed above.
For instance, $\alphaR$ differs by one order of magnitude. %
Interestingly, the massive splitting of $210$~meV observed along $\Gamma$-$M$ in the KTO system is perfectly captured by the fit [Fig.~\ref{fig:Delta-k-Fit}(c)] %
and thus emerges as strong nonlinear $k^3$ response ($\alphaRRR \sim 1.2$~eV~$\AA^{3}$) of the $d_{xy}$ orbitals.

\begin{figure*}
\begin{center}
\includegraphics[width=\textwidth]{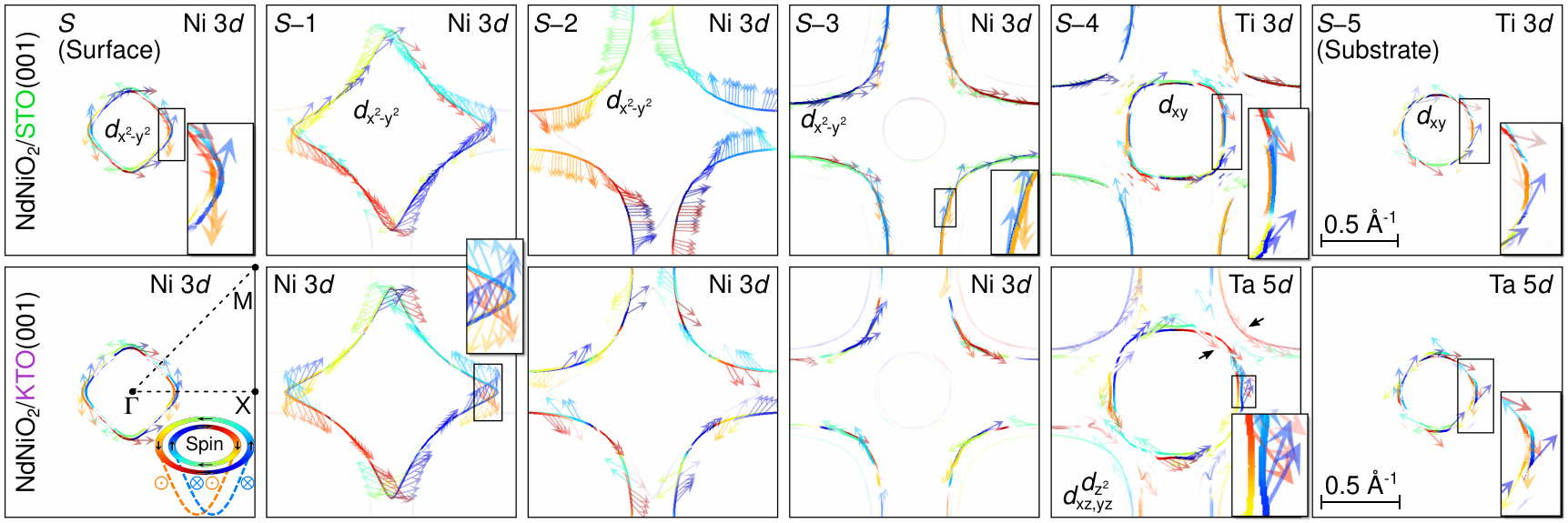}
\caption{\label{fig:FermiSurfaces} Layer-resolved Fermi surface  %
of (NdNiO$_2$)$_4$/STO(001) (top) and (NdNiO$_2$)$_4$/KTO(001) (bottom).
The line intensity represents the respective $d$ orbital character, whereas the color and arrows indicate the basal spin direction (cf.~Fig.~\ref{fig:BandsSOC}).
The reconstructed Ni $3d_{x^2-y^2}$-derived Fermi sheet at the surface
undergoes a Lifshitz transition towards a cuprate-like topology deeper in the film,
accompanied by a realignment of the spin texture.
Spin-orbit interactions generally split each contribution into two distinct sheets of opposite spin. %
An exception are the 'single-band' zones along $\Gamma$-$M$ resulting from the suppression of the counter-clockwise Ta~$5d$ cone at the interface [black arrows; cf.~Fig.~\ref{fig:BandsSOC}(c)].
\vspace{-1em}
}
\end{center}
\end{figure*}

At the NdNiO$_2$/KTO(001) interface, $\alphaR$ reaches $\sim 24$~meV~$\AA$ (Table~\ref{tab:AlphaR}),
which compares favorably to $\sim 20$~meV~$\AA$ for LaAlO$_3$/STO(001) determined from magnetotransport measurements~\cite{LAOSTO-Rashba-CavigliaTriscone:10}. %
Moreover, all Rashba splittings observed here exceed $ \sim 4$~meV reported for LaAlO$_3$/STO(001) \cite{LAOSTO-Rashba-CavigliaTriscone:10}. %
In particular, they are one or two orders of magnitude larger than the superconducting gap, %
which amounts to $2$-$4$~meV in $12$-$25\%$ Sr-doped NdNiO$_2$ films on STO(001) according to scanning tunneling spectroscopy~\cite{Gu-NNO2:20}.
Therefore, spin-orbit effects turn out to be a key ingredient in describing the interactions between the superconducting infinite-layer nickelate film and the 2DEG in the substrate.

\textit{Lifshitz transition and spin texture. --}
Finally, we investigate the Fermi surface and the corresponding spin texture
of NdNiO$_2$/STO(001) and NdNiO$_2$/KTO(001).
Figure~\ref{fig:FermiSurfaces} disentangles the $d$-orbital contributions from subsequent $B$O$_2$ layers,
similar to the layer-resolved band structure in Fig.~\ref{fig:LayerBands}.
Owing to the electrostatic doping, each layer exhibits a distinct sheet of unique shape. %
Spin-orbit interactions generally split each contribution into two sheets of opposite spin,
e.g., clockwise versus counter-clockwise spin rotation,
which is indicative of the underlying Rashba spin-orbit physics~\cite{Kepenekian-Rashba-Dresselhaus-OrganicPerov:15}.

At the surface of the nickelate film ($S$),
we observe a significantly reconstructed Ni-$3d_{x^2-y^2}$-derived sheet of almost circular shape
centered around the $\Gamma$ point.
The reconstruction can be attributed to the electrostatic doping %
and suggests a substantial modification of the the local pairing mechanism (e.g., to $s$-wave~\cite{WuThomale-SurfaceSWave-IL:20}),
which may be key in the interpretation of scanning tunneling spectroscopy results~\cite{Gu-NNO2:20}.
This is of particular relevance since, based on these measurements,
it has recently been argued
that the superconductivity in infinite-layer nickelates is of conventional type~\cite{LiLouie-IL-PhononSC:22}. %
On the other hand, the considerable Ni-O buckling at the surface (cf.~Fig.~\ref{fig:LayerBands})
rather evokes parallels to Fe-based superconductors~\cite{ChubukovHirschfeld-FeSC:15}.

From layer $S-1$ to layer $S-2$, a topological transition of the Fermi sheets to a cuprate-like shape occurs.
This Lifshitz transition~\cite{Lifshitz:60} is accompanied by a reorientation of the spins, which ultimately form a closed loop around the $M$ point ($S-3$). %
The Fermi sheets are degenerate for layer $S-2$, %
whereas layers $S-1$ and $S-3$ present a clear Rashba splitting.

At the interface ($S-4$),
the $d_{xy}$-derived 2DEG sheets centered around the $\Gamma$ point are combined
with $d_{xz,yz}$ and $d_{z^2}$ contributions
centered around the $M$ point.
The enhanced splitting in NdNiO$_2$/KTO(001) is clearly visible (cf.~Table~\ref{tab:AlphaR}).
Interestingly, 'single-band' zones emerge along $\Gamma$-$M$ due to the suppression of the counter-clockwise Ta~$5d$ cone [black arrows in Fig.~\ref{fig:FermiSurfaces}].
In turn, weak $d_{xz,yz}$ signatures appear around the $X$ point that are absent in the STO system.
The overall shape of the Fermi sheet and its spin texture %
is reminiscent of the oxygen-deficient (i.e., $n$-type) STO(001) surface~\cite{AltmeyerValenti-STO-SpinTexture:16}.
Finally, layer $S-5$ presents a quasi-circular Rashba-split 2DEG sheet of exclusive $d_{xy}$ character centered around the $\Gamma$ point.

\textit{Summary. --}
We investigated the impact of spin-orbit interactions in NdNiO$_2$/SrTiO$_3$(001) and NdNiO$_2$/KTaO$_3$(001)
in explicit film geometry.
First-principles simulations unraveled that
the distinct polarity mismatch at the Nd$^{3+}$/(TiO$_2$)$^{0}$ versus Nd$^{3+}$/(TaO$_2$)$^{1+}$ interface
results in the emergence of a two-dimensional electron gas (2DEG) in NdNiO$_2$/KTaO$_3$(001) %
that is twice as pronounced as in NdNiO$_2$/SrTiO$_3$(001)
and involves the occupation of Ta $5d$ conduction-band states, predominantly $d_{xy}$.
Ferroelectric-like displacements of the Ta ions in the substrate act as a fingerprint of this electronic reconstruction.
We identified an anisotropic Rashba spin splitting of the states at the surface and the interface %
that exceeds the size of the superconducting gap,
which establishes spin-orbit effects as crucial ingredient in describing the superconductor--2DEG coupling.
At the KTaO$_3$ interface, the splitting reaches $210$~meV, which is reminiscent of Bi(111) surface states.
The predominantly cubic momentum dependence of the spin splitting exhibited by the substrate Ti and Ta $d_{xy}$ states
contrasts with a strong linear component presented by the Ni $3d_{x^2-y^2}$ orbitals in the film,
which exemplifies the orbital selectivity of the Rashba effect.
Finally, we observed a reconstructed, quasi-circular Ni $3d_{x^2-y^2}$-derived Fermi sheet at the surface owing to the electrostatic doping,
which undergoes a Lifshitz transition towards a cuprate-like topology and a concomitant realignment of its spin texture deeper in the film.
These results suggest surface and interface polarity as promising control parameters to tune spin-orbit physics in infinite-layer nickelate heterostructures. %

\begin{acknowledgments}
This work was funded by the National Science Foundation via grant No.~NSF-DMR-2118718, which is gratefully acknowledged.
The author thanks Armin Sahinovic, Dr.~Okan K\"oksal, Prof.~Rossitza Pentcheva, Prof.~Richard Hennig, and Prof.~Peter Hirschfeld for their support and numerous discussions.
\end{acknowledgments}


\begin{thebibliography}{85}%
\makeatletter
\providecommand \@ifxundefined [1]{%
 \@ifx{#1\undefined}
}%
\providecommand \@ifnum [1]{%
 \ifnum #1\expandafter \@firstoftwo
 \else \expandafter \@secondoftwo
 \fi
}%
\providecommand \@ifx [1]{%
 \ifx #1\expandafter \@firstoftwo
 \else \expandafter \@secondoftwo
 \fi
}%
\providecommand \natexlab [1]{#1}%
\providecommand \enquote  [1]{``#1''}%
\providecommand \bibnamefont  [1]{#1}%
\providecommand \bibfnamefont [1]{#1}%
\providecommand \citenamefont [1]{#1}%
\providecommand \href@noop [0]{\@secondoftwo}%
\providecommand \href [0]{\begingroup \@sanitize@url \@href}%
\providecommand \@href[1]{\@@startlink{#1}\@@href}%
\providecommand \@@href[1]{\endgroup#1\@@endlink}%
\providecommand \@sanitize@url [0]{\catcode `\\12\catcode `\$12\catcode
  `\&12\catcode `\#12\catcode `\^12\catcode `\_12\catcode `\%12\relax}%
\providecommand \@@startlink[1]{}%
\providecommand \@@endlink[0]{}%
\providecommand \url  [0]{\begingroup\@sanitize@url \@url }%
\providecommand \@url [1]{\endgroup\@href {#1}{\urlprefix }}%
\providecommand \urlprefix  [0]{URL }%
\providecommand \Eprint [0]{\href }%
\providecommand \doibase [0]{https://doi.org/}%
\providecommand \selectlanguage [0]{\@gobble}%
\providecommand \bibinfo  [0]{\@secondoftwo}%
\providecommand \bibfield  [0]{\@secondoftwo}%
\providecommand \translation [1]{[#1]}%
\providecommand \BibitemOpen [0]{}%
\providecommand \bibitemStop [0]{}%
\providecommand \bibitemNoStop [0]{.\EOS\space}%
\providecommand \EOS [0]{\spacefactor3000\relax}%
\providecommand \BibitemShut  [1]{\csname bibitem#1\endcsname}%
\let\auto@bib@innerbib\@empty
\bibitem [{\citenamefont {Bychkov}\ and\ \citenamefont
  {Rashba}(1984)}]{BychkovRashba:84}%
  \BibitemOpen
  \bibfield  {author} {\bibinfo {author} {\bibfnamefont {Y.~A.}\ \bibnamefont
  {Bychkov}}\ and\ \bibinfo {author} {\bibfnamefont {E.~I.}\ \bibnamefont
  {Rashba}},\ }\bibfield  {title} {\bibinfo {title} {Oscillatory effects and
  the magnetic susceptibility of carriers in inversion layers},\ }\href
  {https://doi.org/10.1088/0022-3719/17/33/015} {\bibfield  {journal} {\bibinfo
   {journal} {Journal of Physics C: Solid State Physics}\ }\textbf {\bibinfo
  {volume} {17}},\ \bibinfo {pages} {6039} (\bibinfo {year}
  {1984})}\BibitemShut {NoStop}%
\bibitem [{\citenamefont {\ifmmode \check{Z}\else
  \v{Z}\fi{}uti\ifmmode~\acute{c}\else \'{c}\fi{}}\ \emph
  {et~al.}(2004)\citenamefont {\ifmmode \check{Z}\else
  \v{Z}\fi{}uti\ifmmode~\acute{c}\else \'{c}\fi{}}, \citenamefont {Fabian},\
  and\ \citenamefont {Das~Sarma}}]{SpintronicsFabian:04}%
  \BibitemOpen
  \bibfield  {author} {\bibinfo {author} {\bibfnamefont {I.}~\bibnamefont
  {\ifmmode \check{Z}\else \v{Z}\fi{}uti\ifmmode~\acute{c}\else \'{c}\fi{}}},
  \bibinfo {author} {\bibfnamefont {J.}~\bibnamefont {Fabian}},\ and\ \bibinfo
  {author} {\bibfnamefont {S.}~\bibnamefont {Das~Sarma}},\ }\bibfield  {title}
  {\bibinfo {title} {Spintronics: Fundamentals and applications},\ }\href
  {https://doi.org/10.1103/RevModPhys.76.323} {\bibfield  {journal} {\bibinfo
  {journal} {Rev. Mod. Phys.}\ }\textbf {\bibinfo {volume} {76}},\ \bibinfo
  {pages} {323} (\bibinfo {year} {2004})}\BibitemShut {NoStop}%
\bibitem [{\citenamefont {Lutchyn}\ \emph {et~al.}(2010)\citenamefont
  {Lutchyn}, \citenamefont {Sau},\ and\ \citenamefont
  {Das~Sarma}}]{Lutchyn:10}%
  \BibitemOpen
  \bibfield  {author} {\bibinfo {author} {\bibfnamefont {R.~M.}\ \bibnamefont
  {Lutchyn}}, \bibinfo {author} {\bibfnamefont {J.~D.}\ \bibnamefont {Sau}},\
  and\ \bibinfo {author} {\bibfnamefont {S.}~\bibnamefont {Das~Sarma}},\
  }\bibfield  {title} {\bibinfo {title} {Majorana fermions and a topological
  phase transition in semiconductor-superconductor heterostructures},\ }\href
  {https://doi.org/10.1103/PhysRevLett.105.077001} {\bibfield  {journal}
  {\bibinfo  {journal} {Phys. Rev. Lett.}\ }\textbf {\bibinfo {volume} {105}},\
  \bibinfo {pages} {077001} (\bibinfo {year} {2010})}\BibitemShut {NoStop}%
\bibitem [{\citenamefont {Oreg}\ \emph {et~al.}(2010)\citenamefont {Oreg},
  \citenamefont {Refael},\ and\ \citenamefont {von Oppen}}]{Oreg:10}%
  \BibitemOpen
  \bibfield  {author} {\bibinfo {author} {\bibfnamefont {Y.}~\bibnamefont
  {Oreg}}, \bibinfo {author} {\bibfnamefont {G.}~\bibnamefont {Refael}},\ and\
  \bibinfo {author} {\bibfnamefont {F.}~\bibnamefont {von Oppen}},\ }\bibfield
  {title} {\bibinfo {title} {Helical liquids and majorana bound states in
  quantum wires},\ }\href {https://doi.org/10.1103/PhysRevLett.105.177002}
  {\bibfield  {journal} {\bibinfo  {journal} {Phys. Rev. Lett.}\ }\textbf
  {\bibinfo {volume} {105}},\ \bibinfo {pages} {177002} (\bibinfo {year}
  {2010})}\BibitemShut {NoStop}%
\bibitem [{\citenamefont {Lutchyn}\ \emph {et~al.}(2018)\citenamefont
  {Lutchyn}, \citenamefont {Bakkers}, \citenamefont {Kouwenhoven},
  \citenamefont {Krogstrup}, \citenamefont {Marcus},\ and\ \citenamefont
  {Oreg}}]{Lutchyn:18}%
  \BibitemOpen
  \bibfield  {author} {\bibinfo {author} {\bibfnamefont {R.~M.}\ \bibnamefont
  {Lutchyn}}, \bibinfo {author} {\bibfnamefont {E.~P. A.~M.}\ \bibnamefont
  {Bakkers}}, \bibinfo {author} {\bibfnamefont {L.~P.}\ \bibnamefont
  {Kouwenhoven}}, \bibinfo {author} {\bibfnamefont {P.}~\bibnamefont
  {Krogstrup}}, \bibinfo {author} {\bibfnamefont {C.~M.}\ \bibnamefont
  {Marcus}},\ and\ \bibinfo {author} {\bibfnamefont {Y.}~\bibnamefont {Oreg}},\
  }\bibfield  {title} {\bibinfo {title} {Majorana zero modes in
  superconductor--semiconductor heterostructures},\ }\href
  {https://doi.org/10.1038/s41578-018-0003-1} {\bibfield  {journal} {\bibinfo
  {journal} {Nature Reviews Materials}\ }\textbf {\bibinfo {volume} {3}},\
  \bibinfo {pages} {52} (\bibinfo {year} {2018})}\BibitemShut {NoStop}%
\bibitem [{\citenamefont {Lorenz}\ \emph {et~al.}(2016)\citenamefont {Lorenz},
  \citenamefont {Rao}, \citenamefont {Venkatesan}, \citenamefont {Fortunato},
  \citenamefont {Barquinha}, \citenamefont {Branquinho}, \citenamefont
  {Salgueiro}, \citenamefont {Martins}, \citenamefont {Carlos}, \citenamefont
  {Liu}, \citenamefont {Shan}, \citenamefont {Grundmann}, \citenamefont
  {Boschker}, \citenamefont {Mukherjee}, \citenamefont {Priyadarshini},
  \citenamefont {DasGupta}, \citenamefont {Rogers}, \citenamefont {Teherani},
  \citenamefont {Sandana}, \citenamefont {Bove}, \citenamefont {Rietwyk},
  \citenamefont {Zaban}, \citenamefont {Veziridis}, \citenamefont {Weidenkaff},
  \citenamefont {Muralidhar}, \citenamefont {Murakami}, \citenamefont {Abel},
  \citenamefont {Fompeyrine}, \citenamefont {Zuniga-Perez}, \citenamefont
  {Ramesh}, \citenamefont {Spaldin}, \citenamefont {Ostanin}, \citenamefont
  {Borisov}, \citenamefont {Mertig}, \citenamefont {Lazenka}, \citenamefont
  {Srinivasan}, \citenamefont {Prellier}, \citenamefont {Uchida}, \citenamefont
  {Kawasaki}, \citenamefont {Pentcheva}, \citenamefont {Gegenwart},
  \citenamefont {Granozio}, \citenamefont {Fontcuberta},\ and\ \citenamefont
  {Pryds}}]{OxideRoadmap:16}%
  \BibitemOpen
  \bibfield  {author} {\bibinfo {author} {\bibfnamefont {M.}~\bibnamefont
  {Lorenz}}, \bibinfo {author} {\bibfnamefont {M.~S.~R.}\ \bibnamefont {Rao}},
  \bibinfo {author} {\bibfnamefont {T.}~\bibnamefont {Venkatesan}}, \bibinfo
  {author} {\bibfnamefont {E.}~\bibnamefont {Fortunato}}, \bibinfo {author}
  {\bibfnamefont {P.}~\bibnamefont {Barquinha}}, \bibinfo {author}
  {\bibfnamefont {R.}~\bibnamefont {Branquinho}}, \bibinfo {author}
  {\bibfnamefont {D.}~\bibnamefont {Salgueiro}}, \bibinfo {author}
  {\bibfnamefont {R.}~\bibnamefont {Martins}}, \bibinfo {author} {\bibfnamefont
  {E.}~\bibnamefont {Carlos}}, \bibinfo {author} {\bibfnamefont
  {A.}~\bibnamefont {Liu}}, \bibinfo {author} {\bibfnamefont {F.~K.}\
  \bibnamefont {Shan}}, \bibinfo {author} {\bibfnamefont {M.}~\bibnamefont
  {Grundmann}}, \bibinfo {author} {\bibfnamefont {H.}~\bibnamefont {Boschker}},
  \bibinfo {author} {\bibfnamefont {J.}~\bibnamefont {Mukherjee}}, \bibinfo
  {author} {\bibfnamefont {M.}~\bibnamefont {Priyadarshini}}, \bibinfo {author}
  {\bibfnamefont {N.}~\bibnamefont {DasGupta}}, \bibinfo {author}
  {\bibfnamefont {D.~J.}\ \bibnamefont {Rogers}}, \bibinfo {author}
  {\bibfnamefont {F.~H.}\ \bibnamefont {Teherani}}, \bibinfo {author}
  {\bibfnamefont {E.~V.}\ \bibnamefont {Sandana}}, \bibinfo {author}
  {\bibfnamefont {P.}~\bibnamefont {Bove}}, \bibinfo {author} {\bibfnamefont
  {K.}~\bibnamefont {Rietwyk}}, \bibinfo {author} {\bibfnamefont
  {A.}~\bibnamefont {Zaban}}, \bibinfo {author} {\bibfnamefont
  {A.}~\bibnamefont {Veziridis}}, \bibinfo {author} {\bibfnamefont
  {A.}~\bibnamefont {Weidenkaff}}, \bibinfo {author} {\bibfnamefont
  {M.}~\bibnamefont {Muralidhar}}, \bibinfo {author} {\bibfnamefont
  {M.}~\bibnamefont {Murakami}}, \bibinfo {author} {\bibfnamefont
  {S.}~\bibnamefont {Abel}}, \bibinfo {author} {\bibfnamefont {J.}~\bibnamefont
  {Fompeyrine}}, \bibinfo {author} {\bibfnamefont {J.}~\bibnamefont
  {Zuniga-Perez}}, \bibinfo {author} {\bibfnamefont {R.}~\bibnamefont
  {Ramesh}}, \bibinfo {author} {\bibfnamefont {N.~A.}\ \bibnamefont {Spaldin}},
  \bibinfo {author} {\bibfnamefont {S.}~\bibnamefont {Ostanin}}, \bibinfo
  {author} {\bibfnamefont {V.}~\bibnamefont {Borisov}}, \bibinfo {author}
  {\bibfnamefont {I.}~\bibnamefont {Mertig}}, \bibinfo {author} {\bibfnamefont
  {V.}~\bibnamefont {Lazenka}}, \bibinfo {author} {\bibfnamefont
  {G.}~\bibnamefont {Srinivasan}}, \bibinfo {author} {\bibfnamefont
  {W.}~\bibnamefont {Prellier}}, \bibinfo {author} {\bibfnamefont
  {M.}~\bibnamefont {Uchida}}, \bibinfo {author} {\bibfnamefont
  {M.}~\bibnamefont {Kawasaki}}, \bibinfo {author} {\bibfnamefont
  {R.}~\bibnamefont {Pentcheva}}, \bibinfo {author} {\bibfnamefont
  {P.}~\bibnamefont {Gegenwart}}, \bibinfo {author} {\bibfnamefont {F.~M.}\
  \bibnamefont {Granozio}}, \bibinfo {author} {\bibfnamefont {J.}~\bibnamefont
  {Fontcuberta}},\ and\ \bibinfo {author} {\bibfnamefont {N.}~\bibnamefont
  {Pryds}},\ }\bibfield  {title} {\bibinfo {title} {The 2016 oxide electronic
  materials and oxide interfaces roadmap},\ }\href
  {http://stacks.iop.org/0022-3727/49/i=43/a=433001} {\bibfield  {journal}
  {\bibinfo  {journal} {J. Phys. D: Appl. Phys.}\ }\textbf {\bibinfo {volume}
  {49}},\ \bibinfo {pages} {433001} (\bibinfo {year} {2016})}\BibitemShut
  {NoStop}%
\bibitem [{\citenamefont {Ohtomo}\ and\ \citenamefont
  {Hwang}(2004)}]{Ohtomo:2004}%
  \BibitemOpen
  \bibfield  {author} {\bibinfo {author} {\bibfnamefont {A.}~\bibnamefont
  {Ohtomo}}\ and\ \bibinfo {author} {\bibfnamefont {H.~Y.}\ \bibnamefont
  {Hwang}},\ }\bibfield  {title} {\bibinfo {title} {A high-mobility electron
  gas at the {LaAlO$_3$/SrTiO$_3$} heterointerface},\ }\href
  {https://doi.org/10.1038/nature02308} {\bibfield  {journal} {\bibinfo
  {journal} {Nature}\ }\textbf {\bibinfo {volume} {427}},\ \bibinfo {pages}
  {423} (\bibinfo {year} {2004})}\BibitemShut {NoStop}%
\bibitem [{\citenamefont {Nakagawa}\ \emph {et~al.}(2006)\citenamefont
  {Nakagawa}, \citenamefont {Hwang},\ and\ \citenamefont
  {Muller}}]{Nakagawa:06}%
  \BibitemOpen
  \bibfield  {author} {\bibinfo {author} {\bibfnamefont {N.}~\bibnamefont
  {Nakagawa}}, \bibinfo {author} {\bibfnamefont {H.~Y.}\ \bibnamefont
  {Hwang}},\ and\ \bibinfo {author} {\bibfnamefont {D.~A.}\ \bibnamefont
  {Muller}},\ }\bibfield  {title} {\bibinfo {title} {Why some interfaces cannot
  be sharp},\ }\href {https://doi.org/10.1038/nmat1569} {\bibfield  {journal}
  {\bibinfo  {journal} {Nat. Mater.}\ }\textbf {\bibinfo {volume} {5}},\
  \bibinfo {pages} {204} (\bibinfo {year} {2006})}\BibitemShut {NoStop}%
\bibitem [{\citenamefont {Thiel}\ \emph {et~al.}(2006)\citenamefont {Thiel},
  \citenamefont {Hammerl}, \citenamefont {Schmehl}, \citenamefont {Schneider},\
  and\ \citenamefont {Mannhart}}]{Thiel:06}%
  \BibitemOpen
  \bibfield  {author} {\bibinfo {author} {\bibfnamefont {S.}~\bibnamefont
  {Thiel}}, \bibinfo {author} {\bibfnamefont {G.}~\bibnamefont {Hammerl}},
  \bibinfo {author} {\bibfnamefont {A.}~\bibnamefont {Schmehl}}, \bibinfo
  {author} {\bibfnamefont {C.~W.}\ \bibnamefont {Schneider}},\ and\ \bibinfo
  {author} {\bibfnamefont {J.}~\bibnamefont {Mannhart}},\ }\bibfield  {title}
  {\bibinfo {title} {Tunable quasi-two-dimensional electron gases in oxide
  heterostructures},\ }\href {https://doi.org/10.1126/science.1131091}
  {\bibfield  {journal} {\bibinfo  {journal} {Science}\ }\textbf {\bibinfo
  {volume} {313}},\ \bibinfo {pages} {1942} (\bibinfo {year}
  {2006})}\BibitemShut {NoStop}%
\bibitem [{\citenamefont {Pentcheva}\ and\ \citenamefont
  {Pickett}(2009)}]{PentchevaPickett:09}%
  \BibitemOpen
  \bibfield  {author} {\bibinfo {author} {\bibfnamefont {R.}~\bibnamefont
  {Pentcheva}}\ and\ \bibinfo {author} {\bibfnamefont {W.~E.}\ \bibnamefont
  {Pickett}},\ }\bibfield  {title} {\bibinfo {title} {Avoiding the polarization
  catastrophe in {${\mathrm{LaAlO}}_{3}$} overlayers on
  {${\mathrm{SrTiO}}_{3}(001)$} through polar distortion},\ }\href
  {https://doi.org/10.1103/PhysRevLett.102.107602} {\bibfield  {journal}
  {\bibinfo  {journal} {Phys. Rev. Lett.}\ }\textbf {\bibinfo {volume} {102}},\
  \bibinfo {pages} {107602} (\bibinfo {year} {2009})}\BibitemShut {NoStop}%
\bibitem [{\citenamefont {Bell}\ \emph {et~al.}(2009)\citenamefont {Bell},
  \citenamefont {Harashima}, \citenamefont {Kozuka}, \citenamefont {Kim},
  \citenamefont {Kim}, \citenamefont {Hikita},\ and\ \citenamefont
  {Hwang}}]{Bell:09}%
  \BibitemOpen
  \bibfield  {author} {\bibinfo {author} {\bibfnamefont {C.}~\bibnamefont
  {Bell}}, \bibinfo {author} {\bibfnamefont {S.}~\bibnamefont {Harashima}},
  \bibinfo {author} {\bibfnamefont {Y.}~\bibnamefont {Kozuka}}, \bibinfo
  {author} {\bibfnamefont {M.}~\bibnamefont {Kim}}, \bibinfo {author}
  {\bibfnamefont {B.~G.}\ \bibnamefont {Kim}}, \bibinfo {author} {\bibfnamefont
  {Y.}~\bibnamefont {Hikita}},\ and\ \bibinfo {author} {\bibfnamefont {H.~Y.}\
  \bibnamefont {Hwang}},\ }\bibfield  {title} {\bibinfo {title} {Dominant
  mobility modulation by the electric field effect at the
  ${\mathrm{laalo}}_{3}/{\mathrm{srtio}}_{3}$ interface},\ }\href
  {https://doi.org/10.1103/PhysRevLett.103.226802} {\bibfield  {journal}
  {\bibinfo  {journal} {Phys. Rev. Lett.}\ }\textbf {\bibinfo {volume} {103}},\
  \bibinfo {pages} {226802} (\bibinfo {year} {2009})}\BibitemShut {NoStop}%
\bibitem [{\citenamefont {Reyren}\ \emph {et~al.}(2007)\citenamefont {Reyren},
  \citenamefont {Thiel}, \citenamefont {Caviglia}, \citenamefont {Kourkoutis},
  \citenamefont {Hammerl}, \citenamefont {Richter}, \citenamefont {Schneider},
  \citenamefont {Kopp}, \citenamefont {R{\"u}etschi}, \citenamefont {Jaccard},
  \citenamefont {Gabay}, \citenamefont {Muller}, \citenamefont {Triscone},\
  and\ \citenamefont {Mannhart}}]{Reyren:07}%
  \BibitemOpen
  \bibfield  {author} {\bibinfo {author} {\bibfnamefont {N.}~\bibnamefont
  {Reyren}}, \bibinfo {author} {\bibfnamefont {S.}~\bibnamefont {Thiel}},
  \bibinfo {author} {\bibfnamefont {A.~D.}\ \bibnamefont {Caviglia}}, \bibinfo
  {author} {\bibfnamefont {L.~F.}\ \bibnamefont {Kourkoutis}}, \bibinfo
  {author} {\bibfnamefont {G.}~\bibnamefont {Hammerl}}, \bibinfo {author}
  {\bibfnamefont {C.}~\bibnamefont {Richter}}, \bibinfo {author} {\bibfnamefont
  {C.~W.}\ \bibnamefont {Schneider}}, \bibinfo {author} {\bibfnamefont
  {T.}~\bibnamefont {Kopp}}, \bibinfo {author} {\bibfnamefont {A.-S.}\
  \bibnamefont {R{\"u}etschi}}, \bibinfo {author} {\bibfnamefont
  {D.}~\bibnamefont {Jaccard}}, \bibinfo {author} {\bibfnamefont
  {M.}~\bibnamefont {Gabay}}, \bibinfo {author} {\bibfnamefont {D.~A.}\
  \bibnamefont {Muller}}, \bibinfo {author} {\bibfnamefont {J.-M.}\
  \bibnamefont {Triscone}},\ and\ \bibinfo {author} {\bibfnamefont
  {J.}~\bibnamefont {Mannhart}},\ }\bibfield  {title} {\bibinfo {title}
  {Superconducting interfaces between insulating oxides},\ }\href
  {https://doi.org/10.1126/science.1146006} {\bibfield  {journal} {\bibinfo
  {journal} {Science}\ }\textbf {\bibinfo {volume} {317}},\ \bibinfo {pages}
  {1196} (\bibinfo {year} {2007})}\BibitemShut {NoStop}%
\bibitem [{\citenamefont {Caviglia}\ \emph {et~al.}(2010)\citenamefont
  {Caviglia}, \citenamefont {Gabay}, \citenamefont {Gariglio}, \citenamefont
  {Reyren}, \citenamefont {Cancellieri},\ and\ \citenamefont
  {Triscone}}]{LAOSTO-Rashba-CavigliaTriscone:10}%
  \BibitemOpen
  \bibfield  {author} {\bibinfo {author} {\bibfnamefont {A.~D.}\ \bibnamefont
  {Caviglia}}, \bibinfo {author} {\bibfnamefont {M.}~\bibnamefont {Gabay}},
  \bibinfo {author} {\bibfnamefont {S.}~\bibnamefont {Gariglio}}, \bibinfo
  {author} {\bibfnamefont {N.}~\bibnamefont {Reyren}}, \bibinfo {author}
  {\bibfnamefont {C.}~\bibnamefont {Cancellieri}},\ and\ \bibinfo {author}
  {\bibfnamefont {J.-M.}\ \bibnamefont {Triscone}},\ }\bibfield  {title}
  {\bibinfo {title} {Tunable {Rashba} spin-orbit interaction at oxide
  interfaces},\ }\href {https://doi.org/10.1103/PhysRevLett.104.126803}
  {\bibfield  {journal} {\bibinfo  {journal} {Phys. Rev. Lett.}\ }\textbf
  {\bibinfo {volume} {104}},\ \bibinfo {pages} {126803} (\bibinfo {year}
  {2010})}\BibitemShut {NoStop}%
\bibitem [{\citenamefont {Ben~Shalom}\ \emph {et~al.}(2010)\citenamefont
  {Ben~Shalom}, \citenamefont {Sachs}, \citenamefont {Rakhmilevitch},
  \citenamefont {Palevski},\ and\ \citenamefont {Dagan}}]{BenShalom:10}%
  \BibitemOpen
  \bibfield  {author} {\bibinfo {author} {\bibfnamefont {M.}~\bibnamefont
  {Ben~Shalom}}, \bibinfo {author} {\bibfnamefont {M.}~\bibnamefont {Sachs}},
  \bibinfo {author} {\bibfnamefont {D.}~\bibnamefont {Rakhmilevitch}}, \bibinfo
  {author} {\bibfnamefont {A.}~\bibnamefont {Palevski}},\ and\ \bibinfo
  {author} {\bibfnamefont {Y.}~\bibnamefont {Dagan}},\ }\bibfield  {title}
  {\bibinfo {title} {Tuning spin-orbit coupling and superconductivity at the
  {${\mathrm{SrTiO}}_{3}/{\mathrm{LaAlO}}_{3}$} interface: A magnetotransport
  study},\ }\href {https://doi.org/10.1103/PhysRevLett.104.126802} {\bibfield
  {journal} {\bibinfo  {journal} {Phys. Rev. Lett.}\ }\textbf {\bibinfo
  {volume} {104}},\ \bibinfo {pages} {126802} (\bibinfo {year}
  {2010})}\BibitemShut {NoStop}%
\bibitem [{\citenamefont {Dikin}\ \emph {et~al.}(2011)\citenamefont {Dikin},
  \citenamefont {Mehta}, \citenamefont {Bark}, \citenamefont {Folkman},
  \citenamefont {Eom},\ and\ \citenamefont {Chandrasekhar}}]{Dikin:11}%
  \BibitemOpen
  \bibfield  {author} {\bibinfo {author} {\bibfnamefont {D.~A.}\ \bibnamefont
  {Dikin}}, \bibinfo {author} {\bibfnamefont {M.}~\bibnamefont {Mehta}},
  \bibinfo {author} {\bibfnamefont {C.~W.}\ \bibnamefont {Bark}}, \bibinfo
  {author} {\bibfnamefont {C.~M.}\ \bibnamefont {Folkman}}, \bibinfo {author}
  {\bibfnamefont {C.~B.}\ \bibnamefont {Eom}},\ and\ \bibinfo {author}
  {\bibfnamefont {V.}~\bibnamefont {Chandrasekhar}},\ }\bibfield  {title}
  {\bibinfo {title} {Coexistence of superconductivity and ferromagnetism in two
  dimensions},\ }\href {https://doi.org/10.1103/PhysRevLett.107.056802}
  {\bibfield  {journal} {\bibinfo  {journal} {Phys. Rev. Lett.}\ }\textbf
  {\bibinfo {volume} {107}},\ \bibinfo {pages} {056802} (\bibinfo {year}
  {2011})}\BibitemShut {NoStop}%
\bibitem [{\citenamefont {Michaeli}\ \emph {et~al.}(2012)\citenamefont
  {Michaeli}, \citenamefont {Potter},\ and\ \citenamefont {Lee}}]{Michaeli:12}%
  \BibitemOpen
  \bibfield  {author} {\bibinfo {author} {\bibfnamefont {K.}~\bibnamefont
  {Michaeli}}, \bibinfo {author} {\bibfnamefont {A.~C.}\ \bibnamefont
  {Potter}},\ and\ \bibinfo {author} {\bibfnamefont {P.~A.}\ \bibnamefont
  {Lee}},\ }\bibfield  {title} {\bibinfo {title} {Superconducting and
  ferromagnetic phases in ${\mathrm{srtio}}_{3}/{\mathrm{laalo}}_{3}$ oxide
  interface structures: Possibility of finite momentum pairing},\ }\href
  {https://doi.org/10.1103/PhysRevLett.108.117003} {\bibfield  {journal}
  {\bibinfo  {journal} {Phys. Rev. Lett.}\ }\textbf {\bibinfo {volume} {108}},\
  \bibinfo {pages} {117003} (\bibinfo {year} {2012})}\BibitemShut {NoStop}%
\bibitem [{\citenamefont {Li}\ \emph {et~al.}(2019)\citenamefont {Li},
  \citenamefont {Lee}, \citenamefont {Wang}, \citenamefont {Osada},
  \citenamefont {Crossley}, \citenamefont {Lee}, \citenamefont {Cui},
  \citenamefont {Hikita},\ and\ \citenamefont
  {Hwang}}]{Li-Supercond-Inf-NNO-STO:19}%
  \BibitemOpen
  \bibfield  {author} {\bibinfo {author} {\bibfnamefont {D.}~\bibnamefont
  {Li}}, \bibinfo {author} {\bibfnamefont {K.}~\bibnamefont {Lee}}, \bibinfo
  {author} {\bibfnamefont {B.~Y.}\ \bibnamefont {Wang}}, \bibinfo {author}
  {\bibfnamefont {M.}~\bibnamefont {Osada}}, \bibinfo {author} {\bibfnamefont
  {S.}~\bibnamefont {Crossley}}, \bibinfo {author} {\bibfnamefont {H.~R.}\
  \bibnamefont {Lee}}, \bibinfo {author} {\bibfnamefont {Y.}~\bibnamefont
  {Cui}}, \bibinfo {author} {\bibfnamefont {Y.}~\bibnamefont {Hikita}},\ and\
  \bibinfo {author} {\bibfnamefont {H.~Y.}\ \bibnamefont {Hwang}},\ }\bibfield
  {title} {\bibinfo {title} {Superconductivity in an infinite-layer
  nickelate},\ }\href {https://doi.org/10.1038/s41586-019-1496-5} {\bibfield
  {journal} {\bibinfo  {journal} {Nature}\ }\textbf {\bibinfo {volume} {572}},\
  \bibinfo {pages} {624} (\bibinfo {year} {2019})}\BibitemShut {NoStop}%
\bibitem [{\citenamefont {Li}\ \emph {et~al.}(2020{\natexlab{a}})\citenamefont
  {Li}, \citenamefont {Wang}, \citenamefont {Lee}, \citenamefont {Harvey},
  \citenamefont {Osada}, \citenamefont {Goodge}, \citenamefont {Kourkoutis},\
  and\ \citenamefont {Hwang}}]{Li-Supercond-Dome-Inf-NNO-STO:20}%
  \BibitemOpen
  \bibfield  {author} {\bibinfo {author} {\bibfnamefont {D.}~\bibnamefont
  {Li}}, \bibinfo {author} {\bibfnamefont {B.~Y.}\ \bibnamefont {Wang}},
  \bibinfo {author} {\bibfnamefont {K.}~\bibnamefont {Lee}}, \bibinfo {author}
  {\bibfnamefont {S.~P.}\ \bibnamefont {Harvey}}, \bibinfo {author}
  {\bibfnamefont {M.}~\bibnamefont {Osada}}, \bibinfo {author} {\bibfnamefont
  {B.~H.}\ \bibnamefont {Goodge}}, \bibinfo {author} {\bibfnamefont {L.~F.}\
  \bibnamefont {Kourkoutis}},\ and\ \bibinfo {author} {\bibfnamefont {H.~Y.}\
  \bibnamefont {Hwang}},\ }\bibfield  {title} {\bibinfo {title}
  {Superconducting dome in
  {${\mathrm{Nd}}_{1\ensuremath{-}x}{\mathrm{Sr}}_{x}{\mathrm{NiO}}_{2}$}
  infinite layer films},\ }\href
  {https://doi.org/10.1103/PhysRevLett.125.027001} {\bibfield  {journal}
  {\bibinfo  {journal} {Phys. Rev. Lett.}\ }\textbf {\bibinfo {volume} {125}},\
  \bibinfo {pages} {027001} (\bibinfo {year} {2020}{\natexlab{a}})}\BibitemShut
  {NoStop}%
\bibitem [{\citenamefont {Osada}\ \emph {et~al.}(2020)\citenamefont {Osada},
  \citenamefont {Wang}, \citenamefont {Goodge}, \citenamefont {Lee},
  \citenamefont {Yoon}, \citenamefont {Sakuma}, \citenamefont {Li},
  \citenamefont {Miura}, \citenamefont {Kourkoutis},\ and\ \citenamefont
  {Hwang}}]{Osada-PrNiO2-SC:20}%
  \BibitemOpen
  \bibfield  {author} {\bibinfo {author} {\bibfnamefont {M.}~\bibnamefont
  {Osada}}, \bibinfo {author} {\bibfnamefont {B.~Y.}\ \bibnamefont {Wang}},
  \bibinfo {author} {\bibfnamefont {B.~H.}\ \bibnamefont {Goodge}}, \bibinfo
  {author} {\bibfnamefont {K.}~\bibnamefont {Lee}}, \bibinfo {author}
  {\bibfnamefont {H.}~\bibnamefont {Yoon}}, \bibinfo {author} {\bibfnamefont
  {K.}~\bibnamefont {Sakuma}}, \bibinfo {author} {\bibfnamefont
  {D.}~\bibnamefont {Li}}, \bibinfo {author} {\bibfnamefont {M.}~\bibnamefont
  {Miura}}, \bibinfo {author} {\bibfnamefont {L.~F.}\ \bibnamefont
  {Kourkoutis}},\ and\ \bibinfo {author} {\bibfnamefont {H.~Y.}\ \bibnamefont
  {Hwang}},\ }\bibfield  {title} {\bibinfo {title} {A superconducting
  praseodymium nickelate with infinite layer structure},\ }\href
  {https://doi.org/10.1021/acs.nanolett.0c01392} {\bibfield  {journal}
  {\bibinfo  {journal} {Nano Lett.}\ }\textbf {\bibinfo {volume} {20}},\
  \bibinfo {pages} {5735} (\bibinfo {year} {2020})}\BibitemShut {NoStop}%
\bibitem [{\citenamefont {Zeng}\ \emph {et~al.}(2020)\citenamefont {Zeng},
  \citenamefont {Tang}, \citenamefont {Yin}, \citenamefont {Li}, \citenamefont
  {Li}, \citenamefont {Huang}, \citenamefont {Hu}, \citenamefont {Liu},
  \citenamefont {Omar}, \citenamefont {Jani}, \citenamefont {Lim},
  \citenamefont {Han}, \citenamefont {Wan}, \citenamefont {Yang}, \citenamefont
  {Pennycook}, \citenamefont {Wee},\ and\ \citenamefont
  {Ariando}}]{Zeng-Inf-NNO:20}%
  \BibitemOpen
  \bibfield  {author} {\bibinfo {author} {\bibfnamefont {S.}~\bibnamefont
  {Zeng}}, \bibinfo {author} {\bibfnamefont {C.~S.}\ \bibnamefont {Tang}},
  \bibinfo {author} {\bibfnamefont {X.}~\bibnamefont {Yin}}, \bibinfo {author}
  {\bibfnamefont {C.}~\bibnamefont {Li}}, \bibinfo {author} {\bibfnamefont
  {M.}~\bibnamefont {Li}}, \bibinfo {author} {\bibfnamefont {Z.}~\bibnamefont
  {Huang}}, \bibinfo {author} {\bibfnamefont {J.}~\bibnamefont {Hu}}, \bibinfo
  {author} {\bibfnamefont {W.}~\bibnamefont {Liu}}, \bibinfo {author}
  {\bibfnamefont {G.~J.}\ \bibnamefont {Omar}}, \bibinfo {author}
  {\bibfnamefont {H.}~\bibnamefont {Jani}}, \bibinfo {author} {\bibfnamefont
  {Z.~S.}\ \bibnamefont {Lim}}, \bibinfo {author} {\bibfnamefont
  {K.}~\bibnamefont {Han}}, \bibinfo {author} {\bibfnamefont {D.}~\bibnamefont
  {Wan}}, \bibinfo {author} {\bibfnamefont {P.}~\bibnamefont {Yang}}, \bibinfo
  {author} {\bibfnamefont {S.~J.}\ \bibnamefont {Pennycook}}, \bibinfo {author}
  {\bibfnamefont {A.~T.~S.}\ \bibnamefont {Wee}},\ and\ \bibinfo {author}
  {\bibfnamefont {A.}~\bibnamefont {Ariando}},\ }\bibfield  {title} {\bibinfo
  {title} {Phase diagram and superconducting dome of infinite-layer
  {${\mathrm{Nd}}_{1\ensuremath{-}x}{\mathrm{Sr}}_{x}{\mathrm{NiO}}_{2}$} thin
  films},\ }\href {https://doi.org/10.1103/PhysRevLett.125.147003} {\bibfield
  {journal} {\bibinfo  {journal} {Phys. Rev. Lett.}\ }\textbf {\bibinfo
  {volume} {125}},\ \bibinfo {pages} {147003} (\bibinfo {year}
  {2020})}\BibitemShut {NoStop}%
\bibitem [{\citenamefont {Nomura}\ \emph {et~al.}(2019)\citenamefont {Nomura},
  \citenamefont {Hirayama}, \citenamefont {Tadano}, \citenamefont {Yoshimoto},
  \citenamefont {Nakamura},\ and\ \citenamefont {Arita}}]{Nomura-Inf-NNO:19}%
  \BibitemOpen
  \bibfield  {author} {\bibinfo {author} {\bibfnamefont {Y.}~\bibnamefont
  {Nomura}}, \bibinfo {author} {\bibfnamefont {M.}~\bibnamefont {Hirayama}},
  \bibinfo {author} {\bibfnamefont {T.}~\bibnamefont {Tadano}}, \bibinfo
  {author} {\bibfnamefont {Y.}~\bibnamefont {Yoshimoto}}, \bibinfo {author}
  {\bibfnamefont {K.}~\bibnamefont {Nakamura}},\ and\ \bibinfo {author}
  {\bibfnamefont {R.}~\bibnamefont {Arita}},\ }\bibfield  {title} {\bibinfo
  {title} {Formation of a two-dimensional single-component correlated electron
  system and band engineering in the nickelate superconductor
  {${\mathrm{NdNiO}}_{2}$}},\ }\href
  {https://doi.org/10.1103/PhysRevB.100.205138} {\bibfield  {journal} {\bibinfo
   {journal} {Phys. Rev. B}\ }\textbf {\bibinfo {volume} {100}},\ \bibinfo
  {pages} {205138} (\bibinfo {year} {2019})}\BibitemShut {NoStop}%
\bibitem [{\citenamefont {Jiang}\ \emph {et~al.}(2019)\citenamefont {Jiang},
  \citenamefont {Si}, \citenamefont {Liao},\ and\ \citenamefont
  {Zhong}}]{JiangZhong-InfNickelates:19}%
  \BibitemOpen
  \bibfield  {author} {\bibinfo {author} {\bibfnamefont {P.}~\bibnamefont
  {Jiang}}, \bibinfo {author} {\bibfnamefont {L.}~\bibnamefont {Si}}, \bibinfo
  {author} {\bibfnamefont {Z.}~\bibnamefont {Liao}},\ and\ \bibinfo {author}
  {\bibfnamefont {Z.}~\bibnamefont {Zhong}},\ }\bibfield  {title} {\bibinfo
  {title} {Electronic structure of rare-earth infinite-layer {$R$NiO$_{2}$}
  {($R=$~La, Nd)}},\ }\href {https://doi.org/10.1103/PhysRevB.100.201106}
  {\bibfield  {journal} {\bibinfo  {journal} {Phys. Rev. B}\ }\textbf {\bibinfo
  {volume} {100}},\ \bibinfo {pages} {201106} (\bibinfo {year}
  {2019})}\BibitemShut {NoStop}%
\bibitem [{\citenamefont {Sakakibara}\ \emph {et~al.}(2020)\citenamefont
  {Sakakibara}, \citenamefont {Usui}, \citenamefont {Suzuki}, \citenamefont
  {Kotani}, \citenamefont {Aoki},\ and\ \citenamefont
  {Kuroki}}]{Sakakibara:20}%
  \BibitemOpen
  \bibfield  {author} {\bibinfo {author} {\bibfnamefont {H.}~\bibnamefont
  {Sakakibara}}, \bibinfo {author} {\bibfnamefont {H.}~\bibnamefont {Usui}},
  \bibinfo {author} {\bibfnamefont {K.}~\bibnamefont {Suzuki}}, \bibinfo
  {author} {\bibfnamefont {T.}~\bibnamefont {Kotani}}, \bibinfo {author}
  {\bibfnamefont {H.}~\bibnamefont {Aoki}},\ and\ \bibinfo {author}
  {\bibfnamefont {K.}~\bibnamefont {Kuroki}},\ }\bibfield  {title} {\bibinfo
  {title} {Model construction and a possibility of cupratelike pairing in a new
  {$d^9$} nickelate superconductor {(Nd,Sr)NiO$_{2}$}},\ }\href
  {https://doi.org/10.1103/PhysRevLett.125.077003} {\bibfield  {journal}
  {\bibinfo  {journal} {Phys. Rev. Lett.}\ }\textbf {\bibinfo {volume} {125}},\
  \bibinfo {pages} {077003} (\bibinfo {year} {2020})}\BibitemShut {NoStop}%
\bibitem [{\citenamefont {Jiang}\ \emph {et~al.}(2020)\citenamefont {Jiang},
  \citenamefont {Berciu},\ and\ \citenamefont
  {Sawatzky}}]{JiangBerciuSawatzky:19}%
  \BibitemOpen
  \bibfield  {author} {\bibinfo {author} {\bibfnamefont {M.}~\bibnamefont
  {Jiang}}, \bibinfo {author} {\bibfnamefont {M.}~\bibnamefont {Berciu}},\ and\
  \bibinfo {author} {\bibfnamefont {G.~A.}\ \bibnamefont {Sawatzky}},\
  }\bibfield  {title} {\bibinfo {title} {Critical nature of the {Ni} spin state
  in doped {${\mathrm{NdNiO}}_{2}$}},\ }\href
  {https://doi.org/10.1103/PhysRevLett.124.207004} {\bibfield  {journal}
  {\bibinfo  {journal} {Phys. Rev. Lett.}\ }\textbf {\bibinfo {volume} {124}},\
  \bibinfo {pages} {207004} (\bibinfo {year} {2020})}\BibitemShut {NoStop}%
\bibitem [{\citenamefont {Botana}\ and\ \citenamefont
  {Norman}(2020)}]{Botana-Inf-Nickelates:19}%
  \BibitemOpen
  \bibfield  {author} {\bibinfo {author} {\bibfnamefont {A.~S.}\ \bibnamefont
  {Botana}}\ and\ \bibinfo {author} {\bibfnamefont {M.~R.}\ \bibnamefont
  {Norman}},\ }\bibfield  {title} {\bibinfo {title} {Similarities and
  differences between {${\mathrm{LaNiO}}_{2}$ and ${\mathrm{CaCuO}}_{2}$} and
  implications for superconductivity},\ }\href
  {https://doi.org/10.1103/PhysRevX.10.011024} {\bibfield  {journal} {\bibinfo
  {journal} {Phys. Rev. X}\ }\textbf {\bibinfo {volume} {10}},\ \bibinfo
  {pages} {011024} (\bibinfo {year} {2020})}\BibitemShut {NoStop}%
\bibitem [{\citenamefont {Lechermann}(2020)}]{Lechermann-Inf:20}%
  \BibitemOpen
  \bibfield  {author} {\bibinfo {author} {\bibfnamefont {F.}~\bibnamefont
  {Lechermann}},\ }\bibfield  {title} {\bibinfo {title} {Late transition metal
  oxides with infinite-layer structure: Nickelates versus cuprates},\ }\href
  {https://doi.org/10.1103/PhysRevB.101.081110} {\bibfield  {journal} {\bibinfo
   {journal} {Phys. Rev. B}\ }\textbf {\bibinfo {volume} {101}},\ \bibinfo
  {pages} {081110} (\bibinfo {year} {2020})}\BibitemShut {NoStop}%
\bibitem [{\citenamefont {Si}\ \emph {et~al.}(2020)\citenamefont {Si},
  \citenamefont {Xiao}, \citenamefont {Kaufmann}, \citenamefont {Tomczak},
  \citenamefont {Lu}, \citenamefont {Zhong},\ and\ \citenamefont
  {Held}}]{Si-Zhonh-Held:InfNNO-Hydrogen:20}%
  \BibitemOpen
  \bibfield  {author} {\bibinfo {author} {\bibfnamefont {L.}~\bibnamefont
  {Si}}, \bibinfo {author} {\bibfnamefont {W.}~\bibnamefont {Xiao}}, \bibinfo
  {author} {\bibfnamefont {J.}~\bibnamefont {Kaufmann}}, \bibinfo {author}
  {\bibfnamefont {J.~M.}\ \bibnamefont {Tomczak}}, \bibinfo {author}
  {\bibfnamefont {Y.}~\bibnamefont {Lu}}, \bibinfo {author} {\bibfnamefont
  {Z.}~\bibnamefont {Zhong}},\ and\ \bibinfo {author} {\bibfnamefont
  {K.}~\bibnamefont {Held}},\ }\bibfield  {title} {\bibinfo {title} {Topotactic
  hydrogen in nickelate superconductors and akin infinite-layer oxides
  {$AB{\mathrm{O}}_{2}$}},\ }\href
  {https://doi.org/10.1103/PhysRevLett.124.166402} {\bibfield  {journal}
  {\bibinfo  {journal} {Phys. Rev. Lett.}\ }\textbf {\bibinfo {volume} {124}},\
  \bibinfo {pages} {166402} (\bibinfo {year} {2020})}\BibitemShut {NoStop}%
\bibitem [{\citenamefont {Wu}\ \emph {et~al.}(2020{\natexlab{a}})\citenamefont
  {Wu}, \citenamefont {Di~Sante}, \citenamefont {Schwemmer}, \citenamefont
  {Hanke}, \citenamefont {Hwang}, \citenamefont {Raghu},\ and\ \citenamefont
  {Thomale}}]{NNO-SC-Thomale:20}%
  \BibitemOpen
  \bibfield  {author} {\bibinfo {author} {\bibfnamefont {X.}~\bibnamefont
  {Wu}}, \bibinfo {author} {\bibfnamefont {D.}~\bibnamefont {Di~Sante}},
  \bibinfo {author} {\bibfnamefont {T.}~\bibnamefont {Schwemmer}}, \bibinfo
  {author} {\bibfnamefont {W.}~\bibnamefont {Hanke}}, \bibinfo {author}
  {\bibfnamefont {H.~Y.}\ \bibnamefont {Hwang}}, \bibinfo {author}
  {\bibfnamefont {S.}~\bibnamefont {Raghu}},\ and\ \bibinfo {author}
  {\bibfnamefont {R.}~\bibnamefont {Thomale}},\ }\bibfield  {title} {\bibinfo
  {title} {Robust {$d_{x^2-y^2}$}-wave superconductivity of infinite-layer
  nickelates},\ }\href {https://doi.org/10.1103/PhysRevB.101.060504} {\bibfield
   {journal} {\bibinfo  {journal} {Phys. Rev. B}\ }\textbf {\bibinfo {volume}
  {101}},\ \bibinfo {pages} {060504} (\bibinfo {year}
  {2020}{\natexlab{a}})}\BibitemShut {NoStop}%
\bibitem [{\citenamefont {Gu}\ \emph {et~al.}(2020)\citenamefont {Gu},
  \citenamefont {Li}, \citenamefont {Wan}, \citenamefont {Li}, \citenamefont
  {Guo}, \citenamefont {Yang}, \citenamefont {Li}, \citenamefont {Zhu},
  \citenamefont {Pan}, \citenamefont {Nie},\ and\ \citenamefont
  {Wen}}]{Gu-NNO2:20}%
  \BibitemOpen
  \bibfield  {author} {\bibinfo {author} {\bibfnamefont {Q.}~\bibnamefont
  {Gu}}, \bibinfo {author} {\bibfnamefont {Y.}~\bibnamefont {Li}}, \bibinfo
  {author} {\bibfnamefont {S.}~\bibnamefont {Wan}}, \bibinfo {author}
  {\bibfnamefont {H.}~\bibnamefont {Li}}, \bibinfo {author} {\bibfnamefont
  {W.}~\bibnamefont {Guo}}, \bibinfo {author} {\bibfnamefont {H.}~\bibnamefont
  {Yang}}, \bibinfo {author} {\bibfnamefont {Q.}~\bibnamefont {Li}}, \bibinfo
  {author} {\bibfnamefont {X.}~\bibnamefont {Zhu}}, \bibinfo {author}
  {\bibfnamefont {X.}~\bibnamefont {Pan}}, \bibinfo {author} {\bibfnamefont
  {Y.}~\bibnamefont {Nie}},\ and\ \bibinfo {author} {\bibfnamefont {H.-H.}\
  \bibnamefont {Wen}},\ }\bibfield  {title} {\bibinfo {title} {Single particle
  tunneling spectrum of superconducting {Nd$_{1-x}$Sr$_x$NiO$_2$} thin films},\
  }\href {https://doi.org/10.1038/s41467-020-19908-1} {\bibfield  {journal}
  {\bibinfo  {journal} {Nat. Commun.}\ }\textbf {\bibinfo {volume} {11}},\
  \bibinfo {pages} {6027} (\bibinfo {year} {2020})}\BibitemShut {NoStop}%
\bibitem [{\citenamefont {Lu}\ \emph {et~al.}(2021)\citenamefont {Lu},
  \citenamefont {Rossi}, \citenamefont {Nag}, \citenamefont {Osada},
  \citenamefont {Li}, \citenamefont {Lee}, \citenamefont {Wang}, \citenamefont
  {Garcia-Fernandez}, \citenamefont {Agrestini}, \citenamefont {Shen},
  \citenamefont {Been}, \citenamefont {Moritz}, \citenamefont {Devereaux},
  \citenamefont {Zaanen}, \citenamefont {Hwang}, \citenamefont {Zhou},\ and\
  \citenamefont {Lee}}]{Lu-MagExNdNiO2:21}%
  \BibitemOpen
  \bibfield  {author} {\bibinfo {author} {\bibfnamefont {H.}~\bibnamefont
  {Lu}}, \bibinfo {author} {\bibfnamefont {M.}~\bibnamefont {Rossi}}, \bibinfo
  {author} {\bibfnamefont {A.}~\bibnamefont {Nag}}, \bibinfo {author}
  {\bibfnamefont {M.}~\bibnamefont {Osada}}, \bibinfo {author} {\bibfnamefont
  {D.~F.}\ \bibnamefont {Li}}, \bibinfo {author} {\bibfnamefont
  {K.}~\bibnamefont {Lee}}, \bibinfo {author} {\bibfnamefont {B.~Y.}\
  \bibnamefont {Wang}}, \bibinfo {author} {\bibfnamefont {M.}~\bibnamefont
  {Garcia-Fernandez}}, \bibinfo {author} {\bibfnamefont {S.}~\bibnamefont
  {Agrestini}}, \bibinfo {author} {\bibfnamefont {Z.~X.}\ \bibnamefont {Shen}},
  \bibinfo {author} {\bibfnamefont {E.~M.}\ \bibnamefont {Been}}, \bibinfo
  {author} {\bibfnamefont {B.}~\bibnamefont {Moritz}}, \bibinfo {author}
  {\bibfnamefont {T.~P.}\ \bibnamefont {Devereaux}}, \bibinfo {author}
  {\bibfnamefont {J.}~\bibnamefont {Zaanen}}, \bibinfo {author} {\bibfnamefont
  {H.~Y.}\ \bibnamefont {Hwang}}, \bibinfo {author} {\bibfnamefont {K.-J.}\
  \bibnamefont {Zhou}},\ and\ \bibinfo {author} {\bibfnamefont {W.~S.}\
  \bibnamefont {Lee}},\ }\bibfield  {title} {\bibinfo {title} {Magnetic
  excitations in infinite-layer nickelates},\ }\href
  {https://doi.org/10.1126/science.abd7726} {\bibfield  {journal} {\bibinfo
  {journal} {Science}\ }\textbf {\bibinfo {volume} {373}},\ \bibinfo {pages}
  {213} (\bibinfo {year} {2021})}\BibitemShut {NoStop}%
\bibitem [{\citenamefont {Ortiz}\ \emph {et~al.}(2021)\citenamefont {Ortiz},
  \citenamefont {Menke}, \citenamefont {Misj\'ak}, \citenamefont {Mantadakis},
  \citenamefont {F\"ursich}, \citenamefont {Schierle}, \citenamefont
  {Logvenov}, \citenamefont {Kaiser}, \citenamefont {Keimer}, \citenamefont
  {Hansmann},\ and\ \citenamefont {Benckiser}}]{Ortiz-NNO:21}%
  \BibitemOpen
  \bibfield  {author} {\bibinfo {author} {\bibfnamefont {R.~A.}\ \bibnamefont
  {Ortiz}}, \bibinfo {author} {\bibfnamefont {H.}~\bibnamefont {Menke}},
  \bibinfo {author} {\bibfnamefont {F.}~\bibnamefont {Misj\'ak}}, \bibinfo
  {author} {\bibfnamefont {D.~T.}\ \bibnamefont {Mantadakis}}, \bibinfo
  {author} {\bibfnamefont {K.}~\bibnamefont {F\"ursich}}, \bibinfo {author}
  {\bibfnamefont {E.}~\bibnamefont {Schierle}}, \bibinfo {author}
  {\bibfnamefont {G.}~\bibnamefont {Logvenov}}, \bibinfo {author}
  {\bibfnamefont {U.}~\bibnamefont {Kaiser}}, \bibinfo {author} {\bibfnamefont
  {B.}~\bibnamefont {Keimer}}, \bibinfo {author} {\bibfnamefont
  {P.}~\bibnamefont {Hansmann}},\ and\ \bibinfo {author} {\bibfnamefont
  {E.}~\bibnamefont {Benckiser}},\ }\bibfield  {title} {\bibinfo {title}
  {Superlattice approach to doping infinite-layer nickelates},\ }\href
  {https://doi.org/10.1103/PhysRevB.104.165137} {\bibfield  {journal} {\bibinfo
   {journal} {Phys. Rev. B}\ }\textbf {\bibinfo {volume} {104}},\ \bibinfo
  {pages} {165137} (\bibinfo {year} {2021})}\BibitemShut {NoStop}%
\bibitem [{\citenamefont {Lechermann}(2021)}]{Lechermann-Inf:21}%
  \BibitemOpen
  \bibfield  {author} {\bibinfo {author} {\bibfnamefont {F.}~\bibnamefont
  {Lechermann}},\ }\bibfield  {title} {\bibinfo {title} {Doping-dependent
  character and possible magnetic ordering of {${\mathrm{NdNiO}}_{2}$}},\
  }\href {https://doi.org/10.1103/PhysRevMaterials.5.044803} {\bibfield
  {journal} {\bibinfo  {journal} {Phys. Rev. Mater.}\ }\textbf {\bibinfo
  {volume} {5}},\ \bibinfo {pages} {044803} (\bibinfo {year}
  {2021})}\BibitemShut {NoStop}%
\bibitem [{\citenamefont {Sahinovic}\ and\ \citenamefont
  {Geisler}(2021)}]{SahinovicGeisler:21}%
  \BibitemOpen
  \bibfield  {author} {\bibinfo {author} {\bibfnamefont {A.}~\bibnamefont
  {Sahinovic}}\ and\ \bibinfo {author} {\bibfnamefont {B.}~\bibnamefont
  {Geisler}},\ }\bibfield  {title} {\bibinfo {title} {Active learning and
  element-embedding approach in neural networks for infinite-layer versus
  perovskite oxides},\ }\href
  {https://doi.org/10.1103/PhysRevResearch.3.L042022} {\bibfield  {journal}
  {\bibinfo  {journal} {Phys. Rev. Research}\ }\textbf {\bibinfo {volume}
  {3}},\ \bibinfo {pages} {L042022} (\bibinfo {year} {2021})}\BibitemShut
  {NoStop}%
\bibitem [{\citenamefont {Wang}\ \emph {et~al.}(2021)\citenamefont {Wang},
  \citenamefont {Li}, \citenamefont {Goodge}, \citenamefont {Lee},
  \citenamefont {Osada}, \citenamefont {Harvey}, \citenamefont {Kourkoutis},
  \citenamefont {Beasley},\ and\ \citenamefont {Hwang}}]{Wang-IL-Pauli:21}%
  \BibitemOpen
  \bibfield  {author} {\bibinfo {author} {\bibfnamefont {B.~Y.}\ \bibnamefont
  {Wang}}, \bibinfo {author} {\bibfnamefont {D.}~\bibnamefont {Li}}, \bibinfo
  {author} {\bibfnamefont {B.~H.}\ \bibnamefont {Goodge}}, \bibinfo {author}
  {\bibfnamefont {K.}~\bibnamefont {Lee}}, \bibinfo {author} {\bibfnamefont
  {M.}~\bibnamefont {Osada}}, \bibinfo {author} {\bibfnamefont {S.~P.}\
  \bibnamefont {Harvey}}, \bibinfo {author} {\bibfnamefont {L.~F.}\
  \bibnamefont {Kourkoutis}}, \bibinfo {author} {\bibfnamefont {M.~R.}\
  \bibnamefont {Beasley}},\ and\ \bibinfo {author} {\bibfnamefont {H.~Y.}\
  \bibnamefont {Hwang}},\ }\bibfield  {title} {\bibinfo {title} {Isotropic
  {Pauli-limited} superconductivity in the infinite-layer nickelate
  {Nd$_{0.775}$Sr$_{0.225}$NiO$_2$}},\ }\href
  {https://doi.org/10.1038/s41567-020-01128-5} {\bibfield  {journal} {\bibinfo
  {journal} {Nat. Phys.}\ }\textbf {\bibinfo {volume} {17}},\ \bibinfo {pages}
  {473} (\bibinfo {year} {2021})}\BibitemShut {NoStop}%
\bibitem [{\citenamefont {Kang}\ and\ \citenamefont
  {Kotliar}(2021)}]{Optical-IL-Kotliar:21}%
  \BibitemOpen
  \bibfield  {author} {\bibinfo {author} {\bibfnamefont {C.-J.}\ \bibnamefont
  {Kang}}\ and\ \bibinfo {author} {\bibfnamefont {G.}~\bibnamefont {Kotliar}},\
  }\bibfield  {title} {\bibinfo {title} {Optical properties of the
  infinite-layer
  ${\mathrm{la}}_{1\ensuremath{-}x}{\mathrm{sr}}_{x}{\mathrm{nio}}_{2}$ and
  hidden hund's physics},\ }\href
  {https://doi.org/10.1103/PhysRevLett.126.127401} {\bibfield  {journal}
  {\bibinfo  {journal} {Phys. Rev. Lett.}\ }\textbf {\bibinfo {volume} {126}},\
  \bibinfo {pages} {127401} (\bibinfo {year} {2021})}\BibitemShut {NoStop}%
\bibitem [{\citenamefont {Zeng}\ \emph {et~al.}(2022)\citenamefont {Zeng},
  \citenamefont {Yin}, \citenamefont {Li}, \citenamefont {Chow}, \citenamefont
  {Tang}, \citenamefont {Han}, \citenamefont {Huang}, \citenamefont {Cao},
  \citenamefont {Wan}, \citenamefont {Zhang}, \citenamefont {Lim},
  \citenamefont {Diao}, \citenamefont {Yang}, \citenamefont {Wee},
  \citenamefont {Pennycook},\ and\ \citenamefont {Ariando}}]{Zeng-Inf-NNO:22}%
  \BibitemOpen
  \bibfield  {author} {\bibinfo {author} {\bibfnamefont {S.~W.}\ \bibnamefont
  {Zeng}}, \bibinfo {author} {\bibfnamefont {X.~M.}\ \bibnamefont {Yin}},
  \bibinfo {author} {\bibfnamefont {C.~J.}\ \bibnamefont {Li}}, \bibinfo
  {author} {\bibfnamefont {L.~E.}\ \bibnamefont {Chow}}, \bibinfo {author}
  {\bibfnamefont {C.~S.}\ \bibnamefont {Tang}}, \bibinfo {author}
  {\bibfnamefont {K.}~\bibnamefont {Han}}, \bibinfo {author} {\bibfnamefont
  {Z.}~\bibnamefont {Huang}}, \bibinfo {author} {\bibfnamefont
  {Y.}~\bibnamefont {Cao}}, \bibinfo {author} {\bibfnamefont {D.~Y.}\
  \bibnamefont {Wan}}, \bibinfo {author} {\bibfnamefont {Z.~T.}\ \bibnamefont
  {Zhang}}, \bibinfo {author} {\bibfnamefont {Z.~S.}\ \bibnamefont {Lim}},
  \bibinfo {author} {\bibfnamefont {C.~Z.}\ \bibnamefont {Diao}}, \bibinfo
  {author} {\bibfnamefont {P.}~\bibnamefont {Yang}}, \bibinfo {author}
  {\bibfnamefont {A.~T.~S.}\ \bibnamefont {Wee}}, \bibinfo {author}
  {\bibfnamefont {S.~J.}\ \bibnamefont {Pennycook}},\ and\ \bibinfo {author}
  {\bibfnamefont {A.}~\bibnamefont {Ariando}},\ }\bibfield  {title} {\bibinfo
  {title} {Observation of perfect diamagnetism and interfacial effect on the
  electronic structures in infinite layer {Nd$_{0.8}$Sr$_{0.2}$NiO$_2$}
  superconductors},\ }\href {https://doi.org/10.1038/s41467-022-28390-w}
  {\bibfield  {journal} {\bibinfo  {journal} {Nature Communications}\ }\textbf
  {\bibinfo {volume} {13}},\ \bibinfo {pages} {743} (\bibinfo {year}
  {2022})}\BibitemShut {NoStop}%
\bibitem [{\citenamefont {Kreisel}\ \emph {et~al.}(2022)\citenamefont
  {Kreisel}, \citenamefont {Andersen}, \citenamefont {R\o{}mer}, \citenamefont
  {Eremin},\ and\ \citenamefont {Lechermann}}]{KreiselLechermann-IL:22}%
  \BibitemOpen
  \bibfield  {author} {\bibinfo {author} {\bibfnamefont {A.}~\bibnamefont
  {Kreisel}}, \bibinfo {author} {\bibfnamefont {B.~M.}\ \bibnamefont
  {Andersen}}, \bibinfo {author} {\bibfnamefont {A.~T.}\ \bibnamefont
  {R\o{}mer}}, \bibinfo {author} {\bibfnamefont {I.~M.}\ \bibnamefont
  {Eremin}},\ and\ \bibinfo {author} {\bibfnamefont {F.}~\bibnamefont
  {Lechermann}},\ }\bibfield  {title} {\bibinfo {title} {Superconducting
  instabilities in strongly correlated infinite-layer nickelates},\ }\href
  {https://doi.org/10.1103/PhysRevLett.129.077002} {\bibfield  {journal}
  {\bibinfo  {journal} {Phys. Rev. Lett.}\ }\textbf {\bibinfo {volume} {129}},\
  \bibinfo {pages} {077002} (\bibinfo {year} {2022})}\BibitemShut {NoStop}%
\bibitem [{\citenamefont {Rossi}\ \emph {et~al.}(2022)\citenamefont {Rossi},
  \citenamefont {Osada}, \citenamefont {Choi}, \citenamefont {Agrestini},
  \citenamefont {Jost}, \citenamefont {Lee}, \citenamefont {Lu}, \citenamefont
  {Wang}, \citenamefont {Lee}, \citenamefont {Nag}, \citenamefont {Chuang},
  \citenamefont {Kuo}, \citenamefont {Lee}, \citenamefont {Moritz},
  \citenamefont {Devereaux}, \citenamefont {Shen}, \citenamefont {Lee},
  \citenamefont {Zhou}, \citenamefont {Hwang},\ and\ \citenamefont
  {Lee}}]{Rossi-IL-CO:22}%
  \BibitemOpen
  \bibfield  {author} {\bibinfo {author} {\bibfnamefont {M.}~\bibnamefont
  {Rossi}}, \bibinfo {author} {\bibfnamefont {M.}~\bibnamefont {Osada}},
  \bibinfo {author} {\bibfnamefont {J.}~\bibnamefont {Choi}}, \bibinfo {author}
  {\bibfnamefont {S.}~\bibnamefont {Agrestini}}, \bibinfo {author}
  {\bibfnamefont {D.}~\bibnamefont {Jost}}, \bibinfo {author} {\bibfnamefont
  {Y.}~\bibnamefont {Lee}}, \bibinfo {author} {\bibfnamefont {H.}~\bibnamefont
  {Lu}}, \bibinfo {author} {\bibfnamefont {B.~Y.}\ \bibnamefont {Wang}},
  \bibinfo {author} {\bibfnamefont {K.}~\bibnamefont {Lee}}, \bibinfo {author}
  {\bibfnamefont {A.}~\bibnamefont {Nag}}, \bibinfo {author} {\bibfnamefont
  {Y.-D.}\ \bibnamefont {Chuang}}, \bibinfo {author} {\bibfnamefont {C.-T.}\
  \bibnamefont {Kuo}}, \bibinfo {author} {\bibfnamefont {S.-J.}\ \bibnamefont
  {Lee}}, \bibinfo {author} {\bibfnamefont {B.}~\bibnamefont {Moritz}},
  \bibinfo {author} {\bibfnamefont {T.~P.}\ \bibnamefont {Devereaux}}, \bibinfo
  {author} {\bibfnamefont {Z.-X.}\ \bibnamefont {Shen}}, \bibinfo {author}
  {\bibfnamefont {J.-S.}\ \bibnamefont {Lee}}, \bibinfo {author} {\bibfnamefont
  {K.-J.}\ \bibnamefont {Zhou}}, \bibinfo {author} {\bibfnamefont {H.~Y.}\
  \bibnamefont {Hwang}},\ and\ \bibinfo {author} {\bibfnamefont {W.-S.}\
  \bibnamefont {Lee}},\ }\bibfield  {title} {\bibinfo {title} {A broken
  translational symmetry state in an infinite-layer nickelate},\ }\href
  {https://doi.org/10.1038/s41567-022-01660-6} {\bibfield  {journal} {\bibinfo
  {journal} {Nature Physics}\ }\textbf {\bibinfo {volume} {18}},\ \bibinfo
  {pages} {869} (\bibinfo {year} {2022})}\BibitemShut {NoStop}%
\bibitem [{\citenamefont {Fowlie}\ \emph {et~al.}(2022)\citenamefont {Fowlie},
  \citenamefont {Hadjimichael}, \citenamefont {Martins}, \citenamefont {Li},
  \citenamefont {Osada}, \citenamefont {Wang}, \citenamefont {Lee},
  \citenamefont {Lee}, \citenamefont {Salman}, \citenamefont {Prokscha},
  \citenamefont {Triscone}, \citenamefont {Hwang},\ and\ \citenamefont
  {Suter}}]{Fowlie-IL-IntrinsicMag:22}%
  \BibitemOpen
  \bibfield  {author} {\bibinfo {author} {\bibfnamefont {J.}~\bibnamefont
  {Fowlie}}, \bibinfo {author} {\bibfnamefont {M.}~\bibnamefont
  {Hadjimichael}}, \bibinfo {author} {\bibfnamefont {M.~M.}\ \bibnamefont
  {Martins}}, \bibinfo {author} {\bibfnamefont {D.}~\bibnamefont {Li}},
  \bibinfo {author} {\bibfnamefont {M.}~\bibnamefont {Osada}}, \bibinfo
  {author} {\bibfnamefont {B.~Y.}\ \bibnamefont {Wang}}, \bibinfo {author}
  {\bibfnamefont {K.}~\bibnamefont {Lee}}, \bibinfo {author} {\bibfnamefont
  {Y.}~\bibnamefont {Lee}}, \bibinfo {author} {\bibfnamefont {Z.}~\bibnamefont
  {Salman}}, \bibinfo {author} {\bibfnamefont {T.}~\bibnamefont {Prokscha}},
  \bibinfo {author} {\bibfnamefont {J.-M.}\ \bibnamefont {Triscone}}, \bibinfo
  {author} {\bibfnamefont {H.~Y.}\ \bibnamefont {Hwang}},\ and\ \bibinfo
  {author} {\bibfnamefont {A.}~\bibnamefont {Suter}},\ }\bibfield  {title}
  {\bibinfo {title} {Intrinsic magnetism in superconducting infinite-layer
  nickelates},\ }\href {https://doi.org/10.1038/s41567-022-01684-y} {\bibfield
  {journal} {\bibinfo  {journal} {Nature Physics}\ }\textbf {\bibinfo {volume}
  {18}},\ \bibinfo {pages} {1043} (\bibinfo {year} {2022})}\BibitemShut
  {NoStop}%
\bibitem [{\citenamefont {Sahinovic}\ and\ \citenamefont
  {Geisler}(2022)}]{SahinovicGeisler:22}%
  \BibitemOpen
  \bibfield  {author} {\bibinfo {author} {\bibfnamefont {A.}~\bibnamefont
  {Sahinovic}}\ and\ \bibinfo {author} {\bibfnamefont {B.}~\bibnamefont
  {Geisler}},\ }\bibfield  {title} {\bibinfo {title} {Quantifying transfer
  learning synergies in infinite-layer and perovskite nitrides, oxides, and
  fluorides},\ }\href {https://doi.org/10.1088/1361-648x/ac5995} {\bibfield
  {journal} {\bibinfo  {journal} {J. Phys.: Condens. Matter}\ }\textbf
  {\bibinfo {volume} {34}},\ \bibinfo {pages} {214003} (\bibinfo {year}
  {2022})}\BibitemShut {NoStop}%
\bibitem [{\citenamefont {Wang}\ \emph {et~al.}(2022)\citenamefont {Wang},
  \citenamefont {Yang}, \citenamefont {Yang}, \citenamefont {Chen},
  \citenamefont {Zhang}, \citenamefont {Zhang}, \citenamefont {Zhu},
  \citenamefont {Uwatoko}, \citenamefont {Gu}, \citenamefont {Dong},
  \citenamefont {Sun}, \citenamefont {Jin},\ and\ \citenamefont
  {Cheng}}]{Wang-Pressure-PNO:22}%
  \BibitemOpen
  \bibfield  {author} {\bibinfo {author} {\bibfnamefont {N.~N.}\ \bibnamefont
  {Wang}}, \bibinfo {author} {\bibfnamefont {M.~W.}\ \bibnamefont {Yang}},
  \bibinfo {author} {\bibfnamefont {Z.}~\bibnamefont {Yang}}, \bibinfo {author}
  {\bibfnamefont {K.~Y.}\ \bibnamefont {Chen}}, \bibinfo {author}
  {\bibfnamefont {H.}~\bibnamefont {Zhang}}, \bibinfo {author} {\bibfnamefont
  {Q.~H.}\ \bibnamefont {Zhang}}, \bibinfo {author} {\bibfnamefont {Z.~H.}\
  \bibnamefont {Zhu}}, \bibinfo {author} {\bibfnamefont {Y.}~\bibnamefont
  {Uwatoko}}, \bibinfo {author} {\bibfnamefont {L.}~\bibnamefont {Gu}},
  \bibinfo {author} {\bibfnamefont {X.~L.}\ \bibnamefont {Dong}}, \bibinfo
  {author} {\bibfnamefont {J.~P.}\ \bibnamefont {Sun}}, \bibinfo {author}
  {\bibfnamefont {K.~J.}\ \bibnamefont {Jin}},\ and\ \bibinfo {author}
  {\bibfnamefont {J.-G.}\ \bibnamefont {Cheng}},\ }\bibfield  {title} {\bibinfo
  {title} {Pressure-induced monotonic enhancement of tc to over 30{\thinspace}k
  in superconducting {Pr$_{0.82}$Sr$_{0.18}$NiO$_2$} thin films},\ }\href
  {https://doi.org/10.1038/s41467-022-32065-x} {\bibfield  {journal} {\bibinfo
  {journal} {Nat. Commun.}\ }\textbf {\bibinfo {volume} {13}},\ \bibinfo
  {pages} {4367} (\bibinfo {year} {2022})}\BibitemShut {NoStop}%
\bibitem [{\citenamefont {Pan}\ \emph {et~al.}(2022)\citenamefont {Pan},
  \citenamefont {Ferenc~Segedin}, \citenamefont {LaBollita}, \citenamefont
  {Song}, \citenamefont {Nica}, \citenamefont {Goodge}, \citenamefont {Pierce},
  \citenamefont {Doyle}, \citenamefont {Novakov}, \citenamefont
  {C{\'o}rdova~Carrizales}, \citenamefont {N'Diaye}, \citenamefont {Shafer},
  \citenamefont {Paik}, \citenamefont {Heron}, \citenamefont {Mason},
  \citenamefont {Yacoby}, \citenamefont {Kourkoutis}, \citenamefont {Erten},
  \citenamefont {Brooks}, \citenamefont {Botana},\ and\ \citenamefont
  {Mundy}}]{Pan-ILSC:22}%
  \BibitemOpen
  \bibfield  {author} {\bibinfo {author} {\bibfnamefont {G.~A.}\ \bibnamefont
  {Pan}}, \bibinfo {author} {\bibfnamefont {D.}~\bibnamefont {Ferenc~Segedin}},
  \bibinfo {author} {\bibfnamefont {H.}~\bibnamefont {LaBollita}}, \bibinfo
  {author} {\bibfnamefont {Q.}~\bibnamefont {Song}}, \bibinfo {author}
  {\bibfnamefont {E.~M.}\ \bibnamefont {Nica}}, \bibinfo {author}
  {\bibfnamefont {B.~H.}\ \bibnamefont {Goodge}}, \bibinfo {author}
  {\bibfnamefont {A.~T.}\ \bibnamefont {Pierce}}, \bibinfo {author}
  {\bibfnamefont {S.}~\bibnamefont {Doyle}}, \bibinfo {author} {\bibfnamefont
  {S.}~\bibnamefont {Novakov}}, \bibinfo {author} {\bibfnamefont
  {D.}~\bibnamefont {C{\'o}rdova~Carrizales}}, \bibinfo {author} {\bibfnamefont
  {A.~T.}\ \bibnamefont {N'Diaye}}, \bibinfo {author} {\bibfnamefont
  {P.}~\bibnamefont {Shafer}}, \bibinfo {author} {\bibfnamefont
  {H.}~\bibnamefont {Paik}}, \bibinfo {author} {\bibfnamefont {J.~T.}\
  \bibnamefont {Heron}}, \bibinfo {author} {\bibfnamefont {J.~A.}\ \bibnamefont
  {Mason}}, \bibinfo {author} {\bibfnamefont {A.}~\bibnamefont {Yacoby}},
  \bibinfo {author} {\bibfnamefont {L.~F.}\ \bibnamefont {Kourkoutis}},
  \bibinfo {author} {\bibfnamefont {O.}~\bibnamefont {Erten}}, \bibinfo
  {author} {\bibfnamefont {C.~M.}\ \bibnamefont {Brooks}}, \bibinfo {author}
  {\bibfnamefont {A.~S.}\ \bibnamefont {Botana}},\ and\ \bibinfo {author}
  {\bibfnamefont {J.~A.}\ \bibnamefont {Mundy}},\ }\bibfield  {title} {\bibinfo
  {title} {Superconductivity in a quintuple-layer square-planar nickelate},\
  }\href {https://doi.org/10.1038/s41563-021-01142-9} {\bibfield  {journal}
  {\bibinfo  {journal} {Nat. Mater.}\ }\textbf {\bibinfo {volume} {21}},\
  \bibinfo {pages} {160} (\bibinfo {year} {2022})}\BibitemShut {NoStop}%
\bibitem [{\citenamefont {Goodge}\ \emph {et~al.}(2022)\citenamefont {Goodge},
  \citenamefont {Geisler}, \citenamefont {Lee}, \citenamefont {Osada},
  \citenamefont {Wang}, \citenamefont {Li}, \citenamefont {Hwang},
  \citenamefont {Pentcheva},\ and\ \citenamefont
  {Kourkoutis}}]{GoodgeGeisler-NNO-IF:22}%
  \BibitemOpen
  \bibfield  {author} {\bibinfo {author} {\bibfnamefont {B.~H.}\ \bibnamefont
  {Goodge}}, \bibinfo {author} {\bibfnamefont {B.}~\bibnamefont {Geisler}},
  \bibinfo {author} {\bibfnamefont {K.}~\bibnamefont {Lee}}, \bibinfo {author}
  {\bibfnamefont {M.}~\bibnamefont {Osada}}, \bibinfo {author} {\bibfnamefont
  {B.~Y.}\ \bibnamefont {Wang}}, \bibinfo {author} {\bibfnamefont
  {D.}~\bibnamefont {Li}}, \bibinfo {author} {\bibfnamefont {H.~Y.}\
  \bibnamefont {Hwang}}, \bibinfo {author} {\bibfnamefont {R.}~\bibnamefont
  {Pentcheva}},\ and\ \bibinfo {author} {\bibfnamefont {L.~F.}\ \bibnamefont
  {Kourkoutis}},\ }\href@noop {} {\bibinfo {title} {Reconstructing the polar
  interface of infinite-layer nickelate thin films}} (\bibinfo {year} {2022}),\
  \Eprint {https://arxiv.org/abs/2201.03613} {arXiv:2201.03613
  [cond-mat.supr-con]} \BibitemShut {NoStop}%
\bibitem [{\citenamefont {Li}\ \emph {et~al.}(2020{\natexlab{b}})\citenamefont
  {Li}, \citenamefont {He}, \citenamefont {Si}, \citenamefont {Zhu},
  \citenamefont {Zhang},\ and\ \citenamefont {Wen}}]{Li-NoSCinBulkDopedNNO:19}%
  \BibitemOpen
  \bibfield  {author} {\bibinfo {author} {\bibfnamefont {Q.}~\bibnamefont
  {Li}}, \bibinfo {author} {\bibfnamefont {C.}~\bibnamefont {He}}, \bibinfo
  {author} {\bibfnamefont {J.}~\bibnamefont {Si}}, \bibinfo {author}
  {\bibfnamefont {X.}~\bibnamefont {Zhu}}, \bibinfo {author} {\bibfnamefont
  {Y.}~\bibnamefont {Zhang}},\ and\ \bibinfo {author} {\bibfnamefont {H.-H.}\
  \bibnamefont {Wen}},\ }\bibfield  {title} {\bibinfo {title} {Absence of
  superconductivity in bulk {Nd$_{1-x}$Sr$_{x}$NiO$_2$}},\ }\href
  {https://doi.org/10.1038/s43246-020-0018-1} {\bibfield  {journal} {\bibinfo
  {journal} {Communications Materials}\ }\textbf {\bibinfo {volume} {1}},\
  \bibinfo {pages} {16} (\bibinfo {year} {2020}{\natexlab{b}})}\BibitemShut
  {NoStop}%
\bibitem [{\citenamefont {Wang}\ \emph {et~al.}(2020)\citenamefont {Wang},
  \citenamefont {Zheng}, \citenamefont {Krivyakina}, \citenamefont {Chmaissem},
  \citenamefont {Lopes}, \citenamefont {Lynn}, \citenamefont {Gallington},
  \citenamefont {Ren}, \citenamefont {Rosenkranz}, \citenamefont {Mitchell},\
  and\ \citenamefont {Phelan}}]{Wang-NoSCinBulkDopedNNO:20}%
  \BibitemOpen
  \bibfield  {author} {\bibinfo {author} {\bibfnamefont {B.-X.}\ \bibnamefont
  {Wang}}, \bibinfo {author} {\bibfnamefont {H.}~\bibnamefont {Zheng}},
  \bibinfo {author} {\bibfnamefont {E.}~\bibnamefont {Krivyakina}}, \bibinfo
  {author} {\bibfnamefont {O.}~\bibnamefont {Chmaissem}}, \bibinfo {author}
  {\bibfnamefont {P.~P.}\ \bibnamefont {Lopes}}, \bibinfo {author}
  {\bibfnamefont {J.~W.}\ \bibnamefont {Lynn}}, \bibinfo {author}
  {\bibfnamefont {L.~C.}\ \bibnamefont {Gallington}}, \bibinfo {author}
  {\bibfnamefont {Y.}~\bibnamefont {Ren}}, \bibinfo {author} {\bibfnamefont
  {S.}~\bibnamefont {Rosenkranz}}, \bibinfo {author} {\bibfnamefont {J.~F.}\
  \bibnamefont {Mitchell}},\ and\ \bibinfo {author} {\bibfnamefont
  {D.}~\bibnamefont {Phelan}},\ }\bibfield  {title} {\bibinfo {title}
  {Synthesis and characterization of bulk
  {${\mathrm{Nd}}_{1\ensuremath{-}x}{\mathrm{Sr}}_{x}\mathrm{Ni}{\mathrm{O}}_{2}$}
  and
  {${\mathrm{Nd}}_{1\ensuremath{-}x}{\mathrm{Sr}}_{x}\mathrm{Ni}{\mathrm{O}}_{3}$}},\
  }\href {https://doi.org/10.1103/PhysRevMaterials.4.084409} {\bibfield
  {journal} {\bibinfo  {journal} {Phys. Rev. Materials}\ }\textbf {\bibinfo
  {volume} {4}},\ \bibinfo {pages} {084409} (\bibinfo {year}
  {2020})}\BibitemShut {NoStop}%
\bibitem [{\citenamefont {Geisler}\ and\ \citenamefont
  {Pentcheva}(2020)}]{GeislerPentcheva-InfNNO:20}%
  \BibitemOpen
  \bibfield  {author} {\bibinfo {author} {\bibfnamefont {B.}~\bibnamefont
  {Geisler}}\ and\ \bibinfo {author} {\bibfnamefont {R.}~\bibnamefont
  {Pentcheva}},\ }\bibfield  {title} {\bibinfo {title} {Fundamental difference
  in the electronic reconstruction of infinite-layer versus perovskite
  neodymium nickelate films on {${\mathrm{SrTiO}}_{3}$(001)}},\ }\href
  {https://doi.org/10.1103/PhysRevB.102.020502} {\bibfield  {journal} {\bibinfo
   {journal} {Phys. Rev. B}\ }\textbf {\bibinfo {volume} {102}},\ \bibinfo
  {pages} {020502(R)} (\bibinfo {year} {2020})}\BibitemShut {NoStop}%
\bibitem [{\citenamefont {Bernardini}\ and\ \citenamefont
  {Cano}(2020)}]{BernardiniCano:20}%
  \BibitemOpen
  \bibfield  {author} {\bibinfo {author} {\bibfnamefont {F.}~\bibnamefont
  {Bernardini}}\ and\ \bibinfo {author} {\bibfnamefont {A.}~\bibnamefont
  {Cano}},\ }\bibfield  {title} {\bibinfo {title} {Stability and electronic
  properties of {LaNiO}$_2$/{SrTiO}$_3$ heterostructures},\ }\href
  {https://doi.org/10.1088/2515-7639/ab9d0f} {\bibfield  {journal} {\bibinfo
  {journal} {Journal of Physics: Materials}\ }\textbf {\bibinfo {volume} {3}},\
  \bibinfo {pages} {03LT01} (\bibinfo {year} {2020})}\BibitemShut {NoStop}%
\bibitem [{\citenamefont {He}\ \emph {et~al.}(2020)\citenamefont {He},
  \citenamefont {Jiang}, \citenamefont {Lu}, \citenamefont {Song},
  \citenamefont {Chen}, \citenamefont {Jin}, \citenamefont {Shui},\ and\
  \citenamefont {Zhong}}]{He-IL:20}%
  \BibitemOpen
  \bibfield  {author} {\bibinfo {author} {\bibfnamefont {R.}~\bibnamefont
  {He}}, \bibinfo {author} {\bibfnamefont {P.}~\bibnamefont {Jiang}}, \bibinfo
  {author} {\bibfnamefont {Y.}~\bibnamefont {Lu}}, \bibinfo {author}
  {\bibfnamefont {Y.}~\bibnamefont {Song}}, \bibinfo {author} {\bibfnamefont
  {M.}~\bibnamefont {Chen}}, \bibinfo {author} {\bibfnamefont {M.}~\bibnamefont
  {Jin}}, \bibinfo {author} {\bibfnamefont {L.}~\bibnamefont {Shui}},\ and\
  \bibinfo {author} {\bibfnamefont {Z.}~\bibnamefont {Zhong}},\ }\bibfield
  {title} {\bibinfo {title} {Polarity-induced electronic and atomic
  reconstruction at {${\mathrm{NdNiO}}_{2}/{\mathrm{SrTiO}}_{3}$} interfaces},\
  }\href {https://doi.org/10.1103/PhysRevB.102.035118} {\bibfield  {journal}
  {\bibinfo  {journal} {Phys. Rev. B}\ }\textbf {\bibinfo {volume} {102}},\
  \bibinfo {pages} {035118} (\bibinfo {year} {2020})}\BibitemShut {NoStop}%
\bibitem [{\citenamefont {Zhang}\ \emph {et~al.}(2020)\citenamefont {Zhang},
  \citenamefont {Lin}, \citenamefont {Hu}, \citenamefont {Moreo}, \citenamefont
  {Dong},\ and\ \citenamefont {Dagotto}}]{Zhang-IL:20}%
  \BibitemOpen
  \bibfield  {author} {\bibinfo {author} {\bibfnamefont {Y.}~\bibnamefont
  {Zhang}}, \bibinfo {author} {\bibfnamefont {L.-F.}\ \bibnamefont {Lin}},
  \bibinfo {author} {\bibfnamefont {W.}~\bibnamefont {Hu}}, \bibinfo {author}
  {\bibfnamefont {A.}~\bibnamefont {Moreo}}, \bibinfo {author} {\bibfnamefont
  {S.}~\bibnamefont {Dong}},\ and\ \bibinfo {author} {\bibfnamefont
  {E.}~\bibnamefont {Dagotto}},\ }\bibfield  {title} {\bibinfo {title}
  {Similarities and differences between nickelate and cuprate films grown on a
  {${\mathrm{SrTiO}}_{3}$} substrate},\ }\href
  {https://doi.org/10.1103/PhysRevB.102.195117} {\bibfield  {journal} {\bibinfo
   {journal} {Phys. Rev. B}\ }\textbf {\bibinfo {volume} {102}},\ \bibinfo
  {pages} {195117} (\bibinfo {year} {2020})}\BibitemShut {NoStop}%
\bibitem [{\citenamefont {Geisler}\ and\ \citenamefont
  {Pentcheva}(2021)}]{GeislerPentcheva-NNOCCOSTO:21}%
  \BibitemOpen
  \bibfield  {author} {\bibinfo {author} {\bibfnamefont {B.}~\bibnamefont
  {Geisler}}\ and\ \bibinfo {author} {\bibfnamefont {R.}~\bibnamefont
  {Pentcheva}},\ }\bibfield  {title} {\bibinfo {title} {Correlated interface
  electron gas in infinite-layer nickelate versus cuprate films on
  {${\mathrm{SrTiO}}_{3}(001)$}},\ }\href
  {https://doi.org/10.1103/PhysRevResearch.3.013261} {\bibfield  {journal}
  {\bibinfo  {journal} {Phys. Rev. Research}\ }\textbf {\bibinfo {volume}
  {3}},\ \bibinfo {pages} {013261} (\bibinfo {year} {2021})}\BibitemShut
  {NoStop}%
\bibitem [{\citenamefont {King}\ \emph {et~al.}(2012)\citenamefont {King},
  \citenamefont {He}, \citenamefont {Eknapakul}, \citenamefont {Buaphet},
  \citenamefont {Mo}, \citenamefont {Kaneko}, \citenamefont {Harashima},
  \citenamefont {Hikita}, \citenamefont {Bahramy}, \citenamefont {Bell},
  \citenamefont {Hussain}, \citenamefont {Tokura}, \citenamefont {Shen},
  \citenamefont {Hwang}, \citenamefont {Baumberger},\ and\ \citenamefont
  {Meevasana}}]{KTaO3-SOC-Hwang:12}%
  \BibitemOpen
  \bibfield  {author} {\bibinfo {author} {\bibfnamefont {P.~D.~C.}\
  \bibnamefont {King}}, \bibinfo {author} {\bibfnamefont {R.~H.}\ \bibnamefont
  {He}}, \bibinfo {author} {\bibfnamefont {T.}~\bibnamefont {Eknapakul}},
  \bibinfo {author} {\bibfnamefont {P.}~\bibnamefont {Buaphet}}, \bibinfo
  {author} {\bibfnamefont {S.-K.}\ \bibnamefont {Mo}}, \bibinfo {author}
  {\bibfnamefont {Y.}~\bibnamefont {Kaneko}}, \bibinfo {author} {\bibfnamefont
  {S.}~\bibnamefont {Harashima}}, \bibinfo {author} {\bibfnamefont
  {Y.}~\bibnamefont {Hikita}}, \bibinfo {author} {\bibfnamefont {M.~S.}\
  \bibnamefont {Bahramy}}, \bibinfo {author} {\bibfnamefont {C.}~\bibnamefont
  {Bell}}, \bibinfo {author} {\bibfnamefont {Z.}~\bibnamefont {Hussain}},
  \bibinfo {author} {\bibfnamefont {Y.}~\bibnamefont {Tokura}}, \bibinfo
  {author} {\bibfnamefont {Z.-X.}\ \bibnamefont {Shen}}, \bibinfo {author}
  {\bibfnamefont {H.~Y.}\ \bibnamefont {Hwang}}, \bibinfo {author}
  {\bibfnamefont {F.}~\bibnamefont {Baumberger}},\ and\ \bibinfo {author}
  {\bibfnamefont {W.}~\bibnamefont {Meevasana}},\ }\bibfield  {title} {\bibinfo
  {title} {Subband structure of a two-dimensional electron gas formed at the
  polar surface of the strong spin-orbit perovskite {${\mathrm{KTaO}}_{3}$}},\
  }\href {https://doi.org/10.1103/PhysRevLett.108.117602} {\bibfield  {journal}
  {\bibinfo  {journal} {Phys. Rev. Lett.}\ }\textbf {\bibinfo {volume} {108}},\
  \bibinfo {pages} {117602} (\bibinfo {year} {2012})}\BibitemShut {NoStop}%
\bibitem [{\citenamefont {Liu}\ \emph {et~al.}(2021)\citenamefont {Liu},
  \citenamefont {Yan}, \citenamefont {Jin}, \citenamefont {Ma}, \citenamefont
  {Hsiao}, \citenamefont {Lin}, \citenamefont {Bretz-Sullivan}, \citenamefont
  {Zhou}, \citenamefont {Pearson}, \citenamefont {Fisher}, \citenamefont
  {Jiang}, \citenamefont {Han}, \citenamefont {Zuo}, \citenamefont {Wen},
  \citenamefont {Fong}, \citenamefont {Sun}, \citenamefont {Zhou},\ and\
  \citenamefont {Bhattacharya}}]{Liu-SC-KTO-111-IFs:21}%
  \BibitemOpen
  \bibfield  {author} {\bibinfo {author} {\bibfnamefont {C.}~\bibnamefont
  {Liu}}, \bibinfo {author} {\bibfnamefont {X.}~\bibnamefont {Yan}}, \bibinfo
  {author} {\bibfnamefont {D.}~\bibnamefont {Jin}}, \bibinfo {author}
  {\bibfnamefont {Y.}~\bibnamefont {Ma}}, \bibinfo {author} {\bibfnamefont
  {H.-W.}\ \bibnamefont {Hsiao}}, \bibinfo {author} {\bibfnamefont
  {Y.}~\bibnamefont {Lin}}, \bibinfo {author} {\bibfnamefont {T.~M.}\
  \bibnamefont {Bretz-Sullivan}}, \bibinfo {author} {\bibfnamefont
  {X.}~\bibnamefont {Zhou}}, \bibinfo {author} {\bibfnamefont {J.}~\bibnamefont
  {Pearson}}, \bibinfo {author} {\bibfnamefont {B.}~\bibnamefont {Fisher}},
  \bibinfo {author} {\bibfnamefont {J.~S.}\ \bibnamefont {Jiang}}, \bibinfo
  {author} {\bibfnamefont {W.}~\bibnamefont {Han}}, \bibinfo {author}
  {\bibfnamefont {J.-M.}\ \bibnamefont {Zuo}}, \bibinfo {author} {\bibfnamefont
  {J.}~\bibnamefont {Wen}}, \bibinfo {author} {\bibfnamefont {D.~D.}\
  \bibnamefont {Fong}}, \bibinfo {author} {\bibfnamefont {J.}~\bibnamefont
  {Sun}}, \bibinfo {author} {\bibfnamefont {H.}~\bibnamefont {Zhou}},\ and\
  \bibinfo {author} {\bibfnamefont {A.}~\bibnamefont {Bhattacharya}},\
  }\bibfield  {title} {\bibinfo {title} {Two-dimensional superconductivity and
  anisotropic transport at {KTaO$_3$} (111) interfaces},\ }\href
  {https://doi.org/10.1126/science.aba5511} {\bibfield  {journal} {\bibinfo
  {journal} {Science}\ }\textbf {\bibinfo {volume} {371}},\ \bibinfo {pages}
  {716} (\bibinfo {year} {2021})}\BibitemShut {NoStop}%
\bibitem [{\citenamefont {Chen}\ \emph {et~al.}(2021)\citenamefont {Chen},
  \citenamefont {Liu}, \citenamefont {Sun}, \citenamefont {Chen}, \citenamefont
  {Liu}, \citenamefont {Zhang}, \citenamefont {Li}, \citenamefont {Zhang},
  \citenamefont {Hong}, \citenamefont {Ren}, \citenamefont {Zhang},
  \citenamefont {Tian}, \citenamefont {Zhou}, \citenamefont {Sun},\ and\
  \citenamefont {Xie}}]{Chen-SC-LAOKTO110:21}%
  \BibitemOpen
  \bibfield  {author} {\bibinfo {author} {\bibfnamefont {Z.}~\bibnamefont
  {Chen}}, \bibinfo {author} {\bibfnamefont {Z.}~\bibnamefont {Liu}}, \bibinfo
  {author} {\bibfnamefont {Y.}~\bibnamefont {Sun}}, \bibinfo {author}
  {\bibfnamefont {X.}~\bibnamefont {Chen}}, \bibinfo {author} {\bibfnamefont
  {Y.}~\bibnamefont {Liu}}, \bibinfo {author} {\bibfnamefont {H.}~\bibnamefont
  {Zhang}}, \bibinfo {author} {\bibfnamefont {H.}~\bibnamefont {Li}}, \bibinfo
  {author} {\bibfnamefont {M.}~\bibnamefont {Zhang}}, \bibinfo {author}
  {\bibfnamefont {S.}~\bibnamefont {Hong}}, \bibinfo {author} {\bibfnamefont
  {T.}~\bibnamefont {Ren}}, \bibinfo {author} {\bibfnamefont {C.}~\bibnamefont
  {Zhang}}, \bibinfo {author} {\bibfnamefont {H.}~\bibnamefont {Tian}},
  \bibinfo {author} {\bibfnamefont {Y.}~\bibnamefont {Zhou}}, \bibinfo {author}
  {\bibfnamefont {J.}~\bibnamefont {Sun}},\ and\ \bibinfo {author}
  {\bibfnamefont {Y.}~\bibnamefont {Xie}},\ }\bibfield  {title} {\bibinfo
  {title} {Two-dimensional superconductivity at the
  {${\mathrm{LaAlO}}_{3}/{\mathrm{KTaO}}_{3}(110)$} heterointerface},\ }\href
  {https://doi.org/10.1103/PhysRevLett.126.026802} {\bibfield  {journal}
  {\bibinfo  {journal} {Phys. Rev. Lett.}\ }\textbf {\bibinfo {volume} {126}},\
  \bibinfo {pages} {026802} (\bibinfo {year} {2021})}\BibitemShut {NoStop}%
\bibitem [{\citenamefont {Gupta}\ \emph {et~al.}(2022)\citenamefont {Gupta},
  \citenamefont {Silotia}, \citenamefont {Kumari}, \citenamefont {Dumen},
  \citenamefont {Goyal}, \citenamefont {Tomar}, \citenamefont {Wadehra},
  \citenamefont {Ayyub},\ and\ \citenamefont
  {Chakraverty}}]{Gupta-KTO-Review:22}%
  \BibitemOpen
  \bibfield  {author} {\bibinfo {author} {\bibfnamefont {A.}~\bibnamefont
  {Gupta}}, \bibinfo {author} {\bibfnamefont {H.}~\bibnamefont {Silotia}},
  \bibinfo {author} {\bibfnamefont {A.}~\bibnamefont {Kumari}}, \bibinfo
  {author} {\bibfnamefont {M.}~\bibnamefont {Dumen}}, \bibinfo {author}
  {\bibfnamefont {S.}~\bibnamefont {Goyal}}, \bibinfo {author} {\bibfnamefont
  {R.}~\bibnamefont {Tomar}}, \bibinfo {author} {\bibfnamefont
  {N.}~\bibnamefont {Wadehra}}, \bibinfo {author} {\bibfnamefont
  {P.}~\bibnamefont {Ayyub}},\ and\ \bibinfo {author} {\bibfnamefont
  {S.}~\bibnamefont {Chakraverty}},\ }\bibfield  {title} {\bibinfo {title}
  {{KTaO$_3$} -- the new kid on the spintronics block},\ }\href
  {https://doi.org/https://doi.org/10.1002/adma.202106481} {\bibfield
  {journal} {\bibinfo  {journal} {Advanced Materials}\ }\textbf {\bibinfo
  {volume} {34}},\ \bibinfo {pages} {2106481} (\bibinfo {year}
  {2022})}\BibitemShut {NoStop}%
\bibitem [{\citenamefont {Zou}\ \emph {et~al.}(2015)\citenamefont {Zou},
  \citenamefont {Ismail-Beigi}, \citenamefont {Kisslinger}, \citenamefont
  {Shen}, \citenamefont {Su}, \citenamefont {Walker},\ and\ \citenamefont
  {Ahn}}]{Zou-LTOKTO-001:15}%
  \BibitemOpen
  \bibfield  {author} {\bibinfo {author} {\bibfnamefont {K.}~\bibnamefont
  {Zou}}, \bibinfo {author} {\bibfnamefont {S.}~\bibnamefont {Ismail-Beigi}},
  \bibinfo {author} {\bibfnamefont {K.}~\bibnamefont {Kisslinger}}, \bibinfo
  {author} {\bibfnamefont {X.}~\bibnamefont {Shen}}, \bibinfo {author}
  {\bibfnamefont {D.}~\bibnamefont {Su}}, \bibinfo {author} {\bibfnamefont
  {F.~J.}\ \bibnamefont {Walker}},\ and\ \bibinfo {author} {\bibfnamefont
  {C.~H.}\ \bibnamefont {Ahn}},\ }\bibfield  {title} {\bibinfo {title}
  {{LaTiO$_3$/KTaO$_3$} interfaces: A new two-dimensional electron gas
  system},\ }\href {https://doi.org/10.1063/1.4914310} {\bibfield  {journal}
  {\bibinfo  {journal} {APL Materials}\ }\textbf {\bibinfo {volume} {3}},\
  \bibinfo {pages} {036104} (\bibinfo {year} {2015})}\BibitemShut {NoStop}%
\bibitem [{\citenamefont {Zhang}\ \emph {et~al.}(2019)\citenamefont {Zhang},
  \citenamefont {Yan}, \citenamefont {Zhang}, \citenamefont {Wang},
  \citenamefont {Xiong}, \citenamefont {Zhang}, \citenamefont {Qi},
  \citenamefont {Zhang}, \citenamefont {Han}, \citenamefont {Wu}, \citenamefont
  {Liu}, \citenamefont {Chen}, \citenamefont {Shen},\ and\ \citenamefont
  {Sun}}]{Zhang-amorphLAOKTO-001:19}%
  \BibitemOpen
  \bibfield  {author} {\bibinfo {author} {\bibfnamefont {H.}~\bibnamefont
  {Zhang}}, \bibinfo {author} {\bibfnamefont {X.}~\bibnamefont {Yan}}, \bibinfo
  {author} {\bibfnamefont {X.}~\bibnamefont {Zhang}}, \bibinfo {author}
  {\bibfnamefont {S.}~\bibnamefont {Wang}}, \bibinfo {author} {\bibfnamefont
  {C.}~\bibnamefont {Xiong}}, \bibinfo {author} {\bibfnamefont
  {H.}~\bibnamefont {Zhang}}, \bibinfo {author} {\bibfnamefont
  {S.}~\bibnamefont {Qi}}, \bibinfo {author} {\bibfnamefont {J.}~\bibnamefont
  {Zhang}}, \bibinfo {author} {\bibfnamefont {F.}~\bibnamefont {Han}}, \bibinfo
  {author} {\bibfnamefont {N.}~\bibnamefont {Wu}}, \bibinfo {author}
  {\bibfnamefont {B.}~\bibnamefont {Liu}}, \bibinfo {author} {\bibfnamefont
  {Y.}~\bibnamefont {Chen}}, \bibinfo {author} {\bibfnamefont {B.}~\bibnamefont
  {Shen}},\ and\ \bibinfo {author} {\bibfnamefont {J.}~\bibnamefont {Sun}},\
  }\bibfield  {title} {\bibinfo {title} {Unusual electric and optical tuning of
  {KTaO$_3$}-based two-dimensional electron gases with 5d orbitals},\ }\href
  {https://doi.org/10.1021/acsnano.8b07622} {\bibfield  {journal} {\bibinfo
  {journal} {ACS Nano}\ }\textbf {\bibinfo {volume} {13}},\ \bibinfo {pages}
  {609} (\bibinfo {year} {2019})}\BibitemShut {NoStop}%
\bibitem [{\citenamefont {Wadehra}\ \emph {et~al.}(2020)\citenamefont
  {Wadehra}, \citenamefont {Tomar}, \citenamefont {Varma}, \citenamefont
  {Gopal}, \citenamefont {Singh}, \citenamefont {Dattagupta},\ and\
  \citenamefont {Chakraverty}}]{Wadehra-LVOKTO001:20}%
  \BibitemOpen
  \bibfield  {author} {\bibinfo {author} {\bibfnamefont {N.}~\bibnamefont
  {Wadehra}}, \bibinfo {author} {\bibfnamefont {R.}~\bibnamefont {Tomar}},
  \bibinfo {author} {\bibfnamefont {R.~M.}\ \bibnamefont {Varma}}, \bibinfo
  {author} {\bibfnamefont {R.~K.}\ \bibnamefont {Gopal}}, \bibinfo {author}
  {\bibfnamefont {Y.}~\bibnamefont {Singh}}, \bibinfo {author} {\bibfnamefont
  {S.}~\bibnamefont {Dattagupta}},\ and\ \bibinfo {author} {\bibfnamefont
  {S.}~\bibnamefont {Chakraverty}},\ }\bibfield  {title} {\bibinfo {title}
  {Planar hall effect and anisotropic magnetoresistance in polar-polar
  interface of {LaVO$_3$-KTaO$_3$} with strong spin-orbit coupling},\ }\href
  {https://doi.org/10.1038/s41467-020-14689-z} {\bibfield  {journal} {\bibinfo
  {journal} {Nature Communications}\ }\textbf {\bibinfo {volume} {11}},\
  \bibinfo {pages} {874} (\bibinfo {year} {2020})}\BibitemShut {NoStop}%
\bibitem [{\citenamefont {Kohn}\ and\ \citenamefont {Sham}(1965)}]{KoSh65}%
  \BibitemOpen
  \bibfield  {author} {\bibinfo {author} {\bibfnamefont {W.}~\bibnamefont
  {Kohn}}\ and\ \bibinfo {author} {\bibfnamefont {L.~J.}\ \bibnamefont
  {Sham}},\ }\bibfield  {title} {\bibinfo {title} {Self-consistent equations
  including exchange and correlation effects},\ }\href
  {https://doi.org/10.1103/PhysRev.140.A1133} {\bibfield  {journal} {\bibinfo
  {journal} {Phys. Rev.}\ }\textbf {\bibinfo {volume} {140}},\ \bibinfo {pages}
  {A1133} (\bibinfo {year} {1965})}\BibitemShut {NoStop}%
\bibitem [{\citenamefont {Kresse}\ and\ \citenamefont
  {Joubert}(1999)}]{USPP-PAW:99}%
  \BibitemOpen
  \bibfield  {author} {\bibinfo {author} {\bibfnamefont {G.}~\bibnamefont
  {Kresse}}\ and\ \bibinfo {author} {\bibfnamefont {D.}~\bibnamefont
  {Joubert}},\ }\bibfield  {title} {\bibinfo {title} {From ultrasoft
  pseudopotentials to the projector augmented-wave method},\ }\href
  {https://doi.org/10.1103/PhysRevB.59.1758} {\bibfield  {journal} {\bibinfo
  {journal} {Phys. Rev. B}\ }\textbf {\bibinfo {volume} {59}},\ \bibinfo
  {pages} {1758} (\bibinfo {year} {1999})}\BibitemShut {NoStop}%
\bibitem [{\citenamefont {Bl\"ochl}(1994)}]{PAW:94}%
  \BibitemOpen
  \bibfield  {author} {\bibinfo {author} {\bibfnamefont {P.~E.}\ \bibnamefont
  {Bl\"ochl}},\ }\bibfield  {title} {\bibinfo {title} {Projector augmented-wave
  method},\ }\href {https://doi.org/10.1103/PhysRevB.50.17953} {\bibfield
  {journal} {\bibinfo  {journal} {Phys. Rev. B}\ }\textbf {\bibinfo {volume}
  {50}},\ \bibinfo {pages} {17953} (\bibinfo {year} {1994})}\BibitemShut
  {NoStop}%
\bibitem [{\citenamefont {Perdew}\ \emph {et~al.}(1996)\citenamefont {Perdew},
  \citenamefont {Burke},\ and\ \citenamefont {Ernzerhof}}]{PeBu96}%
  \BibitemOpen
  \bibfield  {author} {\bibinfo {author} {\bibfnamefont {J.~P.}\ \bibnamefont
  {Perdew}}, \bibinfo {author} {\bibfnamefont {K.}~\bibnamefont {Burke}},\ and\
  \bibinfo {author} {\bibfnamefont {M.}~\bibnamefont {Ernzerhof}},\ }\bibfield
  {title} {\bibinfo {title} {Generalized gradient approximation made simple},\
  }\href {https://doi.org/10.1103/PhysRevLett.77.3865} {\bibfield  {journal}
  {\bibinfo  {journal} {Phys. Rev. Lett.}\ }\textbf {\bibinfo {volume} {77}},\
  \bibinfo {pages} {3865} (\bibinfo {year} {1996})}\BibitemShut {NoStop}%
\bibitem [{\citenamefont {Liechtenstein}\ \emph {et~al.}(1995)\citenamefont
  {Liechtenstein}, \citenamefont {Anisimov},\ and\ \citenamefont
  {Zaanen}}]{LiechtensteinAnisimov:95}%
  \BibitemOpen
  \bibfield  {author} {\bibinfo {author} {\bibfnamefont {A.~I.}\ \bibnamefont
  {Liechtenstein}}, \bibinfo {author} {\bibfnamefont {V.~I.}\ \bibnamefont
  {Anisimov}},\ and\ \bibinfo {author} {\bibfnamefont {J.}~\bibnamefont
  {Zaanen}},\ }\bibfield  {title} {\bibinfo {title} {Density-functional theory
  and strong interactions: {Orbital} ordering in {Mott-Hubbard} insulators},\
  }\href {https://doi.org/10.1103/PhysRevB.52.R5467} {\bibfield  {journal}
  {\bibinfo  {journal} {Phys. Rev. B}\ }\textbf {\bibinfo {volume} {52}},\
  \bibinfo {pages} {R5467} (\bibinfo {year} {1995})}\BibitemShut {NoStop}%
\bibitem [{\citenamefont {Dudarev}\ \emph {et~al.}(1998)\citenamefont
  {Dudarev}, \citenamefont {Botton}, \citenamefont {Savrasov}, \citenamefont
  {Humphreys},\ and\ \citenamefont {Sutton}}]{Dudarev:98}%
  \BibitemOpen
  \bibfield  {author} {\bibinfo {author} {\bibfnamefont {S.~L.}\ \bibnamefont
  {Dudarev}}, \bibinfo {author} {\bibfnamefont {G.~A.}\ \bibnamefont {Botton}},
  \bibinfo {author} {\bibfnamefont {S.~Y.}\ \bibnamefont {Savrasov}}, \bibinfo
  {author} {\bibfnamefont {C.~J.}\ \bibnamefont {Humphreys}},\ and\ \bibinfo
  {author} {\bibfnamefont {A.~P.}\ \bibnamefont {Sutton}},\ }\bibfield  {title}
  {\bibinfo {title} {Electron-energy-loss spectra and the structural stability
  of nickel oxide: An {LSDA$+U$} study},\ }\href
  {https://doi.org/10.1103/PhysRevB.57.1505} {\bibfield  {journal} {\bibinfo
  {journal} {Phys. Rev. B}\ }\textbf {\bibinfo {volume} {57}},\ \bibinfo
  {pages} {1505} (\bibinfo {year} {1998})}\BibitemShut {NoStop}%
\bibitem [{\citenamefont {Liu}\ \emph {et~al.}(2013)\citenamefont {Liu},
  \citenamefont {Kargarian}, \citenamefont {Kareev}, \citenamefont {Gray},
  \citenamefont {Ryan}, \citenamefont {Cruz}, \citenamefont {Tahir},
  \citenamefont {Chuang}, \citenamefont {Guo}, \citenamefont {Rondinelli},
  \citenamefont {Freeland}, \citenamefont {Fiete},\ and\ \citenamefont
  {Chakhalian}}]{Liu-NNO:13}%
  \BibitemOpen
  \bibfield  {author} {\bibinfo {author} {\bibfnamefont {J.}~\bibnamefont
  {Liu}}, \bibinfo {author} {\bibfnamefont {M.}~\bibnamefont {Kargarian}},
  \bibinfo {author} {\bibfnamefont {M.}~\bibnamefont {Kareev}}, \bibinfo
  {author} {\bibfnamefont {B.}~\bibnamefont {Gray}}, \bibinfo {author}
  {\bibfnamefont {P.~J.}\ \bibnamefont {Ryan}}, \bibinfo {author}
  {\bibfnamefont {A.}~\bibnamefont {Cruz}}, \bibinfo {author} {\bibfnamefont
  {N.}~\bibnamefont {Tahir}}, \bibinfo {author} {\bibfnamefont {Y.-D.}\
  \bibnamefont {Chuang}}, \bibinfo {author} {\bibfnamefont {J.}~\bibnamefont
  {Guo}}, \bibinfo {author} {\bibfnamefont {J.~M.}\ \bibnamefont {Rondinelli}},
  \bibinfo {author} {\bibfnamefont {J.~W.}\ \bibnamefont {Freeland}}, \bibinfo
  {author} {\bibfnamefont {G.~A.}\ \bibnamefont {Fiete}},\ and\ \bibinfo
  {author} {\bibfnamefont {J.}~\bibnamefont {Chakhalian}},\ }\bibfield  {title}
  {\bibinfo {title} {Heterointerface engineered electronic and magnetic phases
  of {NdNiO$_3$} thin films},\ }\href {https://doi.org/10.1038/ncomms3714}
  {\bibfield  {journal} {\bibinfo  {journal} {Nat. Commun.}\ }\textbf {\bibinfo
  {volume} {4}},\ \bibinfo {pages} {2714} (\bibinfo {year} {2013})}\BibitemShut
  {NoStop}%
\bibitem [{\citenamefont {Guo}\ \emph {et~al.}(2017)\citenamefont {Guo},
  \citenamefont {Gangopadhyay}, \citenamefont {K{\"o}ksal}, \citenamefont
  {Pentcheva},\ and\ \citenamefont {Pickett}}]{Guo:17}%
  \BibitemOpen
  \bibfield  {author} {\bibinfo {author} {\bibfnamefont {H.}~\bibnamefont
  {Guo}}, \bibinfo {author} {\bibfnamefont {S.}~\bibnamefont {Gangopadhyay}},
  \bibinfo {author} {\bibfnamefont {O.}~\bibnamefont {K{\"o}ksal}}, \bibinfo
  {author} {\bibfnamefont {R.}~\bibnamefont {Pentcheva}},\ and\ \bibinfo
  {author} {\bibfnamefont {W.~E.}\ \bibnamefont {Pickett}},\ }\bibfield
  {title} {\bibinfo {title} {Wide gap {Chern} {Mott} insulating phases achieved
  by design},\ }\href {https://doi.org/10.1038/s41535-016-0007-2} {\bibfield
  {journal} {\bibinfo  {journal} {npj Quantum Materials}\ }\textbf {\bibinfo
  {volume} {2}},\ \bibinfo {pages} {4} (\bibinfo {year} {2017})}\BibitemShut
  {NoStop}%
\bibitem [{\citenamefont {Geisler}\ \emph {et~al.}(2017)\citenamefont
  {Geisler}, \citenamefont {Blanca-Romero},\ and\ \citenamefont
  {Pentcheva}}]{Geisler-LNOSTO:17}%
  \BibitemOpen
  \bibfield  {author} {\bibinfo {author} {\bibfnamefont {B.}~\bibnamefont
  {Geisler}}, \bibinfo {author} {\bibfnamefont {A.}~\bibnamefont
  {Blanca-Romero}},\ and\ \bibinfo {author} {\bibfnamefont {R.}~\bibnamefont
  {Pentcheva}},\ }\bibfield  {title} {\bibinfo {title} {Design of $n$- and
  $p$-type oxide thermoelectrics in {LaNiO$_{3}$/SrTiO$_{3}(001)$}
  superlattices exploiting interface polarity},\ }\href
  {https://doi.org/10.1103/PhysRevB.95.125301} {\bibfield  {journal} {\bibinfo
  {journal} {Phys. Rev. B}\ }\textbf {\bibinfo {volume} {95}},\ \bibinfo
  {pages} {125301} (\bibinfo {year} {2017})}\BibitemShut {NoStop}%
\bibitem [{\citenamefont {Wrobel}\ \emph {et~al.}(2018)\citenamefont {Wrobel},
  \citenamefont {Geisler}, \citenamefont {Wang}, \citenamefont {Christiani},
  \citenamefont {Logvenov}, \citenamefont {Bluschke}, \citenamefont {Schierle},
  \citenamefont {van Aken}, \citenamefont {Keimer}, \citenamefont {Pentcheva},\
  and\ \citenamefont {Benckiser}}]{WrobelGeisler:18}%
  \BibitemOpen
  \bibfield  {author} {\bibinfo {author} {\bibfnamefont {F.}~\bibnamefont
  {Wrobel}}, \bibinfo {author} {\bibfnamefont {B.}~\bibnamefont {Geisler}},
  \bibinfo {author} {\bibfnamefont {Y.}~\bibnamefont {Wang}}, \bibinfo {author}
  {\bibfnamefont {G.}~\bibnamefont {Christiani}}, \bibinfo {author}
  {\bibfnamefont {G.}~\bibnamefont {Logvenov}}, \bibinfo {author}
  {\bibfnamefont {M.}~\bibnamefont {Bluschke}}, \bibinfo {author}
  {\bibfnamefont {E.}~\bibnamefont {Schierle}}, \bibinfo {author}
  {\bibfnamefont {P.~A.}\ \bibnamefont {van Aken}}, \bibinfo {author}
  {\bibfnamefont {B.}~\bibnamefont {Keimer}}, \bibinfo {author} {\bibfnamefont
  {R.}~\bibnamefont {Pentcheva}},\ and\ \bibinfo {author} {\bibfnamefont
  {E.}~\bibnamefont {Benckiser}},\ }\bibfield  {title} {\bibinfo {title}
  {Digital modulation of the nickel valence state in a cuprate-nickelate
  heterostructure},\ }\href {https://doi.org/10.1103/PhysRevMaterials.2.035001}
  {\bibfield  {journal} {\bibinfo  {journal} {Phys. Rev. Materials}\ }\textbf
  {\bibinfo {volume} {2}},\ \bibinfo {pages} {035001} (\bibinfo {year}
  {2018})}\BibitemShut {NoStop}%
\bibitem [{\citenamefont {Geisler}\ and\ \citenamefont
  {Pentcheva}(2018)}]{GeislerPentcheva-LNOLAO:18}%
  \BibitemOpen
  \bibfield  {author} {\bibinfo {author} {\bibfnamefont {B.}~\bibnamefont
  {Geisler}}\ and\ \bibinfo {author} {\bibfnamefont {R.}~\bibnamefont
  {Pentcheva}},\ }\bibfield  {title} {\bibinfo {title} {Confinement- and
  strain-induced enhancement of thermoelectric properties in
  {LaNiO$_{3}$}/{LaAlO$_{3}(001)$} superlattices},\ }\href
  {https://doi.org/10.1103/PhysRevMaterials.2.055403} {\bibfield  {journal}
  {\bibinfo  {journal} {Phys. Rev. Materials}\ }\textbf {\bibinfo {volume}
  {2}},\ \bibinfo {pages} {055403} (\bibinfo {year} {2018})}\BibitemShut
  {NoStop}%
\bibitem [{\citenamefont {Geisler}\ and\ \citenamefont
  {Pentcheva}(2019)}]{GeislerPentcheva-LNOLAO-Resonances:19}%
  \BibitemOpen
  \bibfield  {author} {\bibinfo {author} {\bibfnamefont {B.}~\bibnamefont
  {Geisler}}\ and\ \bibinfo {author} {\bibfnamefont {R.}~\bibnamefont
  {Pentcheva}},\ }\bibfield  {title} {\bibinfo {title} {Inducing $n$- and
  $p$-type thermoelectricity in oxide superlattices by strain tuning of
  orbital-selective transport resonances},\ }\href
  {https://doi.org/10.1103/PhysRevApplied.11.044047} {\bibfield  {journal}
  {\bibinfo  {journal} {Phys. Rev. Applied}\ }\textbf {\bibinfo {volume}
  {11}},\ \bibinfo {pages} {044047} (\bibinfo {year} {2019})}\BibitemShut
  {NoStop}%
\bibitem [{\citenamefont {K{\"o}ksal}\ and\ \citenamefont
  {Pentcheva}(2019)}]{Koeksal:19}%
  \BibitemOpen
  \bibfield  {author} {\bibinfo {author} {\bibfnamefont {O.}~\bibnamefont
  {K{\"o}ksal}}\ and\ \bibinfo {author} {\bibfnamefont {R.}~\bibnamefont
  {Pentcheva}},\ }\bibfield  {title} {\bibinfo {title} {Chern and {Z2}
  topological insulating phases in perovskite-derived 4d and 5d oxide buckled
  honeycomb lattices},\ }\href {https://doi.org/10.1038/s41598-019-53125-1}
  {\bibfield  {journal} {\bibinfo  {journal} {Scientific Reports}\ }\textbf
  {\bibinfo {volume} {9}},\ \bibinfo {pages} {17306} (\bibinfo {year}
  {2019})}\BibitemShut {NoStop}%
\bibitem [{\citenamefont {Geisler}\ \emph {et~al.}(2022)\citenamefont
  {Geisler}, \citenamefont {Follmann},\ and\ \citenamefont
  {Pentcheva}}]{Geisler-VO-LNOLAO:22}%
  \BibitemOpen
  \bibfield  {author} {\bibinfo {author} {\bibfnamefont {B.}~\bibnamefont
  {Geisler}}, \bibinfo {author} {\bibfnamefont {S.}~\bibnamefont {Follmann}},\
  and\ \bibinfo {author} {\bibfnamefont {R.}~\bibnamefont {Pentcheva}},\
  }\bibfield  {title} {\bibinfo {title} {Oxygen vacancy formation and
  electronic reconstruction in strained {${\mathrm{LaNiO}}_{3}$} and
  {${\mathrm{LaNiO}}_{3}/{\mathrm{LaAlO}}_{3}$} superlattices},\ }\href
  {https://doi.org/10.1103/PhysRevB.106.155139} {\bibfield  {journal} {\bibinfo
   {journal} {Phys. Rev. B}\ }\textbf {\bibinfo {volume} {106}},\ \bibinfo
  {pages} {155139} (\bibinfo {year} {2022})}\BibitemShut {NoStop}%
\bibitem [{\citenamefont {Monkhorst}\ and\ \citenamefont
  {Pack}(1976)}]{MoPa76}%
  \BibitemOpen
  \bibfield  {author} {\bibinfo {author} {\bibfnamefont {H.~J.}\ \bibnamefont
  {Monkhorst}}\ and\ \bibinfo {author} {\bibfnamefont {J.~D.}\ \bibnamefont
  {Pack}},\ }\bibfield  {title} {\bibinfo {title} {Special points for
  brillouin-zone integrations},\ }\href
  {https://doi.org/10.1103/PhysRevB.13.5188} {\bibfield  {journal} {\bibinfo
  {journal} {Phys. Rev. B}\ }\textbf {\bibinfo {volume} {13}},\ \bibinfo
  {pages} {5188} (\bibinfo {year} {1976})}\BibitemShut {NoStop}%
\bibitem [{\citenamefont {Zhang}\ \emph {et~al.}(2014)\citenamefont {Zhang},
  \citenamefont {Soltan}, \citenamefont {Schmid}, \citenamefont {Habermeier},
  \citenamefont {Keimer},\ and\ \citenamefont {Kaiser}}]{ZhangKeimer:14}%
  \BibitemOpen
  \bibfield  {author} {\bibinfo {author} {\bibfnamefont {Z.}~\bibnamefont
  {Zhang}}, \bibinfo {author} {\bibfnamefont {S.}~\bibnamefont {Soltan}},
  \bibinfo {author} {\bibfnamefont {H.}~\bibnamefont {Schmid}}, \bibinfo
  {author} {\bibfnamefont {H.-U.}\ \bibnamefont {Habermeier}}, \bibinfo
  {author} {\bibfnamefont {B.}~\bibnamefont {Keimer}},\ and\ \bibinfo {author}
  {\bibfnamefont {U.}~\bibnamefont {Kaiser}},\ }\bibfield  {title} {\bibinfo
  {title} {Revealing the atomic and electronic structure of a
  {SrTiO3/LaNiO3/SrTiO3} heterostructure interface},\ }\href
  {https://doi.org/http://dx.doi.org/10.1063/1.4868513} {\bibfield  {journal}
  {\bibinfo  {journal} {J. Appl. Phys.}\ }\textbf {\bibinfo {volume} {115}},\
  \bibinfo {eid} {103519} (\bibinfo {year} {2014})}\BibitemShut {NoStop}%
\bibitem [{\citenamefont {Hwang}\ \emph {et~al.}(2013)\citenamefont {Hwang},
  \citenamefont {Son}, \citenamefont {Zhang}, \citenamefont {Janotti},
  \citenamefont {Van~de Walle},\ and\ \citenamefont {Stemmer}}]{Hwang:13}%
  \BibitemOpen
  \bibfield  {author} {\bibinfo {author} {\bibfnamefont {J.}~\bibnamefont
  {Hwang}}, \bibinfo {author} {\bibfnamefont {J.}~\bibnamefont {Son}}, \bibinfo
  {author} {\bibfnamefont {J.~Y.}\ \bibnamefont {Zhang}}, \bibinfo {author}
  {\bibfnamefont {A.}~\bibnamefont {Janotti}}, \bibinfo {author} {\bibfnamefont
  {C.~G.}\ \bibnamefont {Van~de Walle}},\ and\ \bibinfo {author} {\bibfnamefont
  {S.}~\bibnamefont {Stemmer}},\ }\bibfield  {title} {\bibinfo {title}
  {Structural origins of the properties of rare earth nickelate
  superlattices},\ }\href {https://doi.org/10.1103/PhysRevB.87.060101}
  {\bibfield  {journal} {\bibinfo  {journal} {Phys. Rev. B}\ }\textbf {\bibinfo
  {volume} {87}},\ \bibinfo {pages} {060101} (\bibinfo {year}
  {2013})}\BibitemShut {NoStop}%
\bibitem [{\citenamefont {Wemple}(1965)}]{KTO-Bulk-Wemple:65}%
  \BibitemOpen
  \bibfield  {author} {\bibinfo {author} {\bibfnamefont {S.~H.}\ \bibnamefont
  {Wemple}},\ }\bibfield  {title} {\bibinfo {title} {Some transport properties
  of oxygen-deficient single-crystal potassium tantalate
  (kta${\mathrm{o}}_{3}$)},\ }\href {https://doi.org/10.1103/PhysRev.137.A1575}
  {\bibfield  {journal} {\bibinfo  {journal} {Phys. Rev.}\ }\textbf {\bibinfo
  {volume} {137}},\ \bibinfo {pages} {A1575} (\bibinfo {year}
  {1965})}\BibitemShut {NoStop}%
\bibitem [{\citenamefont {H\"ochli}\ \emph {et~al.}(1977)\citenamefont
  {H\"ochli}, \citenamefont {Weibel},\ and\ \citenamefont
  {Boatner}}]{KTO-noFE:77}%
  \BibitemOpen
  \bibfield  {author} {\bibinfo {author} {\bibfnamefont {U.~T.}\ \bibnamefont
  {H\"ochli}}, \bibinfo {author} {\bibfnamefont {H.~E.}\ \bibnamefont
  {Weibel}},\ and\ \bibinfo {author} {\bibfnamefont {L.~A.}\ \bibnamefont
  {Boatner}},\ }\bibfield  {title} {\bibinfo {title} {Quantum limit of
  ferroelectric phase transitions in
  {$\mathrm{K}{\mathrm{Ta}}_{1\ensuremath{-}x}{\mathrm{Nb}}_{x}{\mathrm{O}}_{3}$}},\
  }\href {https://doi.org/10.1103/PhysRevLett.39.1158} {\bibfield  {journal}
  {\bibinfo  {journal} {Phys. Rev. Lett.}\ }\textbf {\bibinfo {volume} {39}},\
  \bibinfo {pages} {1158} (\bibinfo {year} {1977})}\BibitemShut {NoStop}%
\bibitem [{\citenamefont {Bihlmayer}(2017)}]{Bihlmayer-Topo:17}%
  \BibitemOpen
  \bibfield  {author} {\bibinfo {author} {\bibfnamefont {G.}~\bibnamefont
  {Bihlmayer}},\ }\bibinfo {title} {{B}and structure from first-principles and
  relativistic effects},\ in\ \href {https://juser.fz-juelich.de/record/830524}
  {\emph {\bibinfo {booktitle} {48th IFF Spring School 2017 "Topological Matter
  - Topological Insulators, Skyrmions and Majoranas", ed. S. Bl\"ugel, Y.
  Mokrousov, T. Sch\"apers, Y. Ando}}}\ (\bibinfo  {publisher}
  {Forschungszentrum J\"ulich GmbH Zentralbibliothek},\ \bibinfo {address}
  {J\"ulich},\ \bibinfo {year} {2017})\BibitemShut {NoStop}%
\bibitem [{\citenamefont {Sahinovic}\ \emph {et~al.}()\citenamefont
  {Sahinovic}, \citenamefont {Geisler},\ and\ \citenamefont
  {Pentcheva}}]{SahinovicGeislerPentcheva:23}%
  \BibitemOpen
  \bibfield  {author} {\bibinfo {author} {\bibfnamefont {A.}~\bibnamefont
  {Sahinovic}}, \bibinfo {author} {\bibfnamefont {B.}~\bibnamefont {Geisler}},\
  and\ \bibinfo {author} {\bibfnamefont {R.}~\bibnamefont {Pentcheva}},\
  }\bibfield  {title} {\bibinfo {title} {Nature of the magnetic coupling in
  bulk infinite-layer nickelates versus cuprates},\ }\href@noop {} {\bibinfo
  {journal} {to be published}\ }\BibitemShut {NoStop}%
\bibitem [{\citenamefont {Koroteev}\ \emph {et~al.}(2004)\citenamefont
  {Koroteev}, \citenamefont {Bihlmayer}, \citenamefont {Gayone}, \citenamefont
  {Chulkov}, \citenamefont {Bl\"ugel}, \citenamefont {Echenique},\ and\
  \citenamefont {Hofmann}}]{Bi111-SOC:04}%
  \BibitemOpen
\bibfield  {journal} {  }\bibfield  {author} {\bibinfo {author} {\bibfnamefont
  {Y.~M.}\ \bibnamefont {Koroteev}}, \bibinfo {author} {\bibfnamefont
  {G.}~\bibnamefont {Bihlmayer}}, \bibinfo {author} {\bibfnamefont {J.~E.}\
  \bibnamefont {Gayone}}, \bibinfo {author} {\bibfnamefont {E.~V.}\
  \bibnamefont {Chulkov}}, \bibinfo {author} {\bibfnamefont {S.}~\bibnamefont
  {Bl\"ugel}}, \bibinfo {author} {\bibfnamefont {P.~M.}\ \bibnamefont
  {Echenique}},\ and\ \bibinfo {author} {\bibfnamefont {P.}~\bibnamefont
  {Hofmann}},\ }\bibfield  {title} {\bibinfo {title} {Strong spin-orbit
  splitting on bi surfaces},\ }\href
  {https://doi.org/10.1103/PhysRevLett.93.046403} {\bibfield  {journal}
  {\bibinfo  {journal} {Phys. Rev. Lett.}\ }\textbf {\bibinfo {volume} {93}},\
  \bibinfo {pages} {046403} (\bibinfo {year} {2004})}\BibitemShut {NoStop}%
\bibitem [{\citenamefont {Kepenekian}\ \emph {et~al.}(2015)\citenamefont
  {Kepenekian}, \citenamefont {Robles}, \citenamefont {Katan}, \citenamefont
  {Sapori}, \citenamefont {Pedesseau},\ and\ \citenamefont
  {Even}}]{Kepenekian-Rashba-Dresselhaus-OrganicPerov:15}%
  \BibitemOpen
  \bibfield  {author} {\bibinfo {author} {\bibfnamefont {M.}~\bibnamefont
  {Kepenekian}}, \bibinfo {author} {\bibfnamefont {R.}~\bibnamefont {Robles}},
  \bibinfo {author} {\bibfnamefont {C.}~\bibnamefont {Katan}}, \bibinfo
  {author} {\bibfnamefont {D.}~\bibnamefont {Sapori}}, \bibinfo {author}
  {\bibfnamefont {L.}~\bibnamefont {Pedesseau}},\ and\ \bibinfo {author}
  {\bibfnamefont {J.}~\bibnamefont {Even}},\ }\bibfield  {title} {\bibinfo
  {title} {Rashba and dresselhaus effects in hybrid organic--inorganic
  perovskites: From basics to devices},\ }\href
  {https://doi.org/10.1021/acsnano.5b04409} {\bibfield  {journal} {\bibinfo
  {journal} {ACS Nano}\ }\textbf {\bibinfo {volume} {9}},\ \bibinfo {pages}
  {11557} (\bibinfo {year} {2015})}\BibitemShut {NoStop}%
\bibitem [{\citenamefont {Wu}\ \emph {et~al.}(2020{\natexlab{b}})\citenamefont
  {Wu}, \citenamefont {Jiang}, \citenamefont {Di~Sante}, \citenamefont {Hanke},
  \citenamefont {Schnyder}, \citenamefont {Hu},\ and\ \citenamefont
  {Thomale}}]{WuThomale-SurfaceSWave-IL:20}%
  \BibitemOpen
  \bibfield  {author} {\bibinfo {author} {\bibfnamefont {X.}~\bibnamefont
  {Wu}}, \bibinfo {author} {\bibfnamefont {K.}~\bibnamefont {Jiang}}, \bibinfo
  {author} {\bibfnamefont {D.}~\bibnamefont {Di~Sante}}, \bibinfo {author}
  {\bibfnamefont {W.}~\bibnamefont {Hanke}}, \bibinfo {author} {\bibfnamefont
  {A.~P.}\ \bibnamefont {Schnyder}}, \bibinfo {author} {\bibfnamefont
  {J.}~\bibnamefont {Hu}},\ and\ \bibinfo {author} {\bibfnamefont
  {R.}~\bibnamefont {Thomale}},\ }\href
  {https://doi.org/10.48550/ARXIV.2008.06009} {\bibinfo {title} {Surface
  $s$-wave superconductivity for oxide-terminated infinite-layer nickelates}}
  (\bibinfo {year} {2020}{\natexlab{b}})\BibitemShut {NoStop}%
\bibitem [{\citenamefont {Li}\ and\ \citenamefont
  {Louie}(2022)}]{LiLouie-IL-PhononSC:22}%
  \BibitemOpen
  \bibfield  {author} {\bibinfo {author} {\bibfnamefont {Z.}~\bibnamefont
  {Li}}\ and\ \bibinfo {author} {\bibfnamefont {S.~G.}\ \bibnamefont {Louie}},\
  }\href {https://doi.org/10.48550/ARXIV.2210.12819} {\bibinfo {title} {Two-gap
  superconductivity and decisive role of rare-earth $d$ electrons in
  infinite-layer nickelates}} (\bibinfo {year} {2022})\BibitemShut {NoStop}%
\bibitem [{\citenamefont {Chubukov}\ and\ \citenamefont
  {Hirschfeld}(2015)}]{ChubukovHirschfeld-FeSC:15}%
  \BibitemOpen
  \bibfield  {author} {\bibinfo {author} {\bibfnamefont {A.}~\bibnamefont
  {Chubukov}}\ and\ \bibinfo {author} {\bibfnamefont {P.~J.}\ \bibnamefont
  {Hirschfeld}},\ }\bibfield  {title} {\bibinfo {title} {Iron-based
  superconductors, seven years later},\ }\href
  {https://doi.org/10.1063/PT.3.2818} {\bibfield  {journal} {\bibinfo
  {journal} {Physics Today}\ }\textbf {\bibinfo {volume} {68}},\ \bibinfo
  {pages} {46} (\bibinfo {year} {2015})}\BibitemShut {NoStop}%
\bibitem [{\citenamefont {Lifshitz}(1960)}]{Lifshitz:60}%
  \BibitemOpen
  \bibfield  {author} {\bibinfo {author} {\bibfnamefont {I.~M.}\ \bibnamefont
  {Lifshitz}},\ }\bibfield  {title} {\bibinfo {title} {Anomalies of electron
  characteristics of a metal in the high pressure region},\ }\href@noop {}
  {\bibfield  {journal} {\bibinfo  {journal} {Sov. Phys. JETP}\ }\textbf
  {\bibinfo {volume} {11}},\ \bibinfo {pages} {1130} (\bibinfo {year}
  {1960})}\BibitemShut {NoStop}%
\bibitem [{\citenamefont {Altmeyer}\ \emph {et~al.}(2016)\citenamefont
  {Altmeyer}, \citenamefont {Jeschke}, \citenamefont {Hijano-Cubelos},
  \citenamefont {Martins}, \citenamefont {Lechermann}, \citenamefont
  {Koepernik}, \citenamefont {Santander-Syro}, \citenamefont {Rozenberg},
  \citenamefont {Valent\'{\i}},\ and\ \citenamefont
  {Gabay}}]{AltmeyerValenti-STO-SpinTexture:16}%
  \BibitemOpen
  \bibfield  {author} {\bibinfo {author} {\bibfnamefont {M.}~\bibnamefont
  {Altmeyer}}, \bibinfo {author} {\bibfnamefont {H.~O.}\ \bibnamefont
  {Jeschke}}, \bibinfo {author} {\bibfnamefont {O.}~\bibnamefont
  {Hijano-Cubelos}}, \bibinfo {author} {\bibfnamefont {C.}~\bibnamefont
  {Martins}}, \bibinfo {author} {\bibfnamefont {F.}~\bibnamefont {Lechermann}},
  \bibinfo {author} {\bibfnamefont {K.}~\bibnamefont {Koepernik}}, \bibinfo
  {author} {\bibfnamefont {A.~F.}\ \bibnamefont {Santander-Syro}}, \bibinfo
  {author} {\bibfnamefont {M.~J.}\ \bibnamefont {Rozenberg}}, \bibinfo {author}
  {\bibfnamefont {R.}~\bibnamefont {Valent\'{\i}}},\ and\ \bibinfo {author}
  {\bibfnamefont {M.}~\bibnamefont {Gabay}},\ }\bibfield  {title} {\bibinfo
  {title} {Magnetism, spin texture, and in-gap states: {Atomic} specialization
  at the surface of oxygen-deficient {${\mathrm{SrTiO}}_{3}$}},\ }\href
  {https://doi.org/10.1103/PhysRevLett.116.157203} {\bibfield  {journal}
  {\bibinfo  {journal} {Phys. Rev. Lett.}\ }\textbf {\bibinfo {volume} {116}},\
  \bibinfo {pages} {157203} (\bibinfo {year} {2016})}\BibitemShut {NoStop}%
\end{thebibliography}
\end{document}